\documentclass{article}

\usepackage{amsmath}
\usepackage{amsfonts}
\usepackage{hyperref}
\usepackage{arxiv}
\usepackage{xspace}
\usepackage{multicol}
\usepackage{multirow}
\usepackage{graphicx}
\usepackage{subcaption}
\usepackage{array}
\usepackage{tcolorbox}
\usepackage[bottom]{footmisc}
\usepackage{adjustbox}
\usepackage{enumitem}
\usepackage{booktabs}
\usepackage{tabularx}
\usepackage{CJKutf8}
\usepackage[ruled, vlined, linesnumbered]{algorithm2e}

\newtcolorbox{rqbox}[1][]{
  colback=gray!5,
  colframe=gray!50,
  boxrule=0.5pt,
  arc=2pt,
  left=3pt, right=3pt,
  top=2pt, bottom=2pt,
  before upper={\textbf{#1: }},
}

\newcommand{\name}{\textsc{Foreword}\xspace} %

\newcommand{\lohsv}{LOHS$_{var}$\xspace\newcommand{\fan}{failure@$25\%$\xspace}}

\newcommand{\toke}{TOK$_{ent}$\xspace}
\newcommand{\tokp}{TOK$_{prob}$\xspace}

\newcommand{\odd}{\texttt{ODD-ADD}\xspace}
\newcommand{\gh}{\texttt{GH-TYPO}\xspace}
\newcommand{\json}{\texttt{JSON-FIX}\xspace}
\newcommand{\model}{\texttt{MODEL-EX}\xspace}
\newcommand{\pos}{\texttt{POS-TAG}\xspace}
\newcommand{\spell}{\texttt{SPELL-CHK}\xspace}
\newcommand{\syn}{\texttt{SYN-BUG}\xspace}
\newcommand{\topic}{\texttt{TOPIC-CLS}\xspace}
\newcommand{\mbpp}{\texttt{MBPP}\xspace}
\newcommand{\live}{\texttt{LCB}\xspace}

\AtBeginDocument{%
  }

\title{\name: Pre-Generation Prediction and Characterisation of LLM Test Failures using the Jacobian Lens}

\author{
Somin Kim\\
KAIST\\
Republic of Korea\\
\texttt{somin.kim@kaist.ac.kr}
\And
Robert Feldt\\
Chalmers University of Technology\\
Sweden\\
\texttt{robert.feldt@chalmers.se}
\And
Shin Yoo\\
KAIST\\
Republic of Korea\\
\texttt{shin.yoo@kaist.ac.kr}
}

\begin{document}
\renewcommand{\shorttitle}{\name: Pre-Generation Prediction and Characterisation of LLM Test Failures using the Jacobian Lens}

\maketitle
\begin{abstract}

Testing LLM-based software is expensive: each input needs several sampled  responses, each judged by a costly oracle. Further, every prompt or model  change calls for re-testing. Pre-generation adequacy measures cut this cost by  ranking inputs by failure risk based on the model's internal state. However, the state of the art, Clotho, models only passing reference inputs, and its  hidden-state representation says nothing about why inputs fail. We present  \name, a framework built on two ideas: learn from both passing and failing  references, and represent inputs through the Jacobian lens (J-lens), an  interpretability technique that reads intermediate activations out as scores over vocabulary tokens. \name pairs these representations with two estimators: a sparse  logistic-regression probe or dual Gaussian Mixture Models (dual GMMs), which also yield a reference density (to flag inputs the references do not cover) and  failing-class components (to group similar failures). On ten tasks and three open-weight LLMs, all four combinations of representations and estimators outperform Clotho on average, both for prediction of failure (ROC-AUC improves by 0.14 on average and by up to 0.31) and test prioritisation (15\% more failures among the 100 top-ranked inputs on average, up to 33\%). Prioritisation obtained using small models also transfers better to proprietary models (on average 22\% and up to 41\% more failures in the first quarter of prioritised inputs). Importantly, our qualitative case study shows that \name supports semantic characterisation of failures by pointing to ambiguities in labels or task specifications. Combined, \name can support testers for prioritising inputs, checking support for a prediction, and characterising specific failures, all without generating output.
\end{abstract}


\section{Introduction}

Large Language Models (LLMs) now power software in a growing number of domains,
from education~\cite{Chu2025aa} and health care~\cite{Qiu2024aa} to software
development itself~\cite{Agarwal2026aa}. Such software must be tested
thoroughly, but LLM testing is expensive. LLMs can react to the same input
differently each time, so a reliable aggregate verdict on one input may take several
iterations, each judged by an oracle that is sometimes itself an
LLM~\cite{Tan2025aa}. Every change to a prompt, model, or configuration
can also require testing again. Testers would therefore like to predict which inputs are likely to fail, understand what failing inputs have in common, and assess when such predictions are supported. Shared failure patterns could help guide fixes 
and help choose tests that cover different kinds of failure. Yet current LLM testing 
tools give little guidance even on which inputs to run first~\cite{dobslaw2026taxonomy}.

Recently, Clotho~\cite{yoon2026clotho}, a pre-generation test adequacy measure, was proposed to estimate the likelihood of failure before an output is generated. The key insight is that output uncertainty can be approximated from the LLM's internal latent space: if a new input's hidden-state representation is an outlier relative to inputs known to produce correct outputs, it is more likely to fail. 
Empirical results show that the ranking lets testers execute inputs likely to fail first, without generating any outputs for the remaining ones.
However, Clotho neither characterises shared failure patterns nor distinguishes familiar failures from inputs unlike its references.

We present \name, a framework that extends Clotho in two ways: (1) learning from both passing and failing reference inputs, and (2) representing inputs through readouts from the Jacobian lens (J-lens), a mechanistic interpretability technique~\cite{gurnee2026verbalizable}. Learning from both outcomes allows us to fit a predictor that distinguishes passing from failing inputs. The J-lens expresses intermediate LLM activations as vocabulary-token scores that reveal concepts the model has available for verbalisation. Unlike Clotho's opaque hidden-state vectors, these scores correspond to readable tokens, offering a basis for describing individual inputs and groups of failures.

\name represents inputs through vocabulary-aligned J-lens Readout Scores (JRS) and supports two failure estimators that learn from both passing and failing reference inputs. The dual Gaussian Mixture Model (GMM) estimator extends Clotho by fitting separate distributions to passing and failing references. Together, the two GMMs yield a failure score, a reference-density signal for flagging weakly supported inputs, and component groups whose token profiles can be inspected separately. Alternatively, a simpler Logistic Regression (LR) estimator can learn the boundary between passing and failing inputs focusing more directly on failure prediction.

Across ten tasks, including two Python program-synthesis benchmarks with executable tests, our evaluation of \name separates representation choice from estimator choice by pairing both estimators with JRS and Clotho's Last-token Input Hidden States (LIHS).
All four configurations improve average ROC-AUC over
Clotho; LR performs best overall. J-lens features provide readable token
profiles with similar average discrimination to LIHS, although differences
vary by task. In our topic-classification case study, component profiles describe separate failure groups, whereas the inspected global LR weight rankings mix several concepts.

Our contributions are as follows:
\begin{itemize}

\item \name, a pre-generation test adequacy framework that learns from
passing and failing references using J-lens features, with LR and dual
GMMs as complementary estimators.

\item An empirical evaluation across ten tasks and six open-weight and
three proprietary LLMs, comparing against Clotho and post-generation
uncertainty metrics and assessing prediction, prioritisation, reference support, transfer, and cost.

\item A qualitative case study examining how J-lens token profiles
characterise failure groups, including groups sharing the same label
confusion.

\end{itemize}

The replication package of \name is publicly available~\cite{replication}.

\section{Background}
\label{sec:background}

This section presents background information relevant to our work.

\subsection{Test Adequacy for Large Language Models}
\label{sec:background-adequacy}

Test adequacy measures how well a given test input, or a set of inputs, checks 
the target System Under Test (SUT). A good adequacy metric is expected to be
correlated with fault detection capability: an input with a higher adequacy 
value should be more likely to reveal unexpected behaviour. Such metrics can be 
used to assess the quality of existing test inputs~\cite{Zhu1997aa}, or to 
prioritise inputs with the aim of revealing unexpected behaviour as early as 
possible~\cite{Elbaum:2001lr}. 

Test adequacy is crucial for efficient testing of LLM-based systems, as their 
test oracles tend to be expensive, be it human judgement or 
LLM-as-judges~\cite{Tan2025aa}. A reliable test adequacy metric will prioritise
inputs according to their likelihood of failing (i.e., producing undesired 
responses), allowing the test efforts to be focused on those problematic 
inputs. 
However, many of the traditional adequacy criteria are white-box, code-based 
metrics, such as structural coverage~\cite{Zhu1997aa}. Consequently, they do 
not directly apply to Large Language Models (LLMs) that are generative models 
with no interpretable structural units.

To mitigate the high oracle cost, confidence metrics have been widely adopted. 
If the LLM response to a given test input has low confidence, it is
more likely that the response is not correct. Various confidence metrics 
have been studied in this context, including token-level likelihood and 
entropy~\cite{Manakul2023aa}, and the semantic-level 
entropy~\cite{Farquhar2024aa}. While the uncertainty measures can be used to 
prioritise test inputs, their common limitation is that uncertainty becomes 
available only after the full response generation, highlighting the fact that 
it is a metric for outputs, not inputs. 

Clotho is a recently proposed pre-generation test adequacy 
metric~\cite{yoon2026clotho}. It approximates the uncertainty with 
out-of-distribution measures at pre-generation stage: the reference 
distribution is obtained by fitting a Gaussian Mixture Model (GMM) using a 
\emph{reference set} of passing input with respect to a specific task, 
iteratively sampled with user labelling. The test adequacy value of an unseen 
input is the negative log likelihood given by the trained GMM. Since the GMM is
fitted with passing inputs, the adequacy score is correlated to the likelihood
of the input generating correct output, and therefore can be used to prioritise 
inputs. Notably, Clotho computes its test adequacy pre-generation, based on the Last-token
Input Hidden State (LIHS), i.e., the hidden state of the last input token at
a layer two thirds of the way through the Transformer decoder. This
representation is known to contain rich semantic information about the given
context~\cite{Azaria2023aa,Zhou2025aa}.

\subsection{Jacobian Lens}
\label{sec:background-jlens}

Hidden states such as LIHS are informative but not human-readable. Lens techniques address this by decoding intermediate activations into the model's vocabulary. In a decoder-only Transformer, each layer reads from and writes to a residual stream, whose value after layer $\ell$ at token position $p$ we denote $a_{\ell,p}$. 
After the last layer, a final normalisation ($\operatorname{norm}$) and the
unembedding matrix ($W_U$) convert the residual stream into next-token logits. The logit lens~\cite{nostalgebraist2020logitlens} applies $\operatorname{norm}$ and $W_U$ directly to intermediate activations. However, this assumes that intermediate layers encode information in the same way as the final layer, which holds only for the last few layers.

The Jacobian lens (J-lens)~\cite{gurnee2026verbalizable} addresses this by first mapping activations into the final layer's representation space, using the average Jacobian $J_\ell=\mathbb E_{x,p,p'\geq p}[\partial a_{\mathrm{final},p'}(x)/\partial a_{\ell,p}(x)]$,
where $x$ ranges over prompts from a generic text corpus, $p$ over source positions, and $p'$ over the current and all later positions. The readout of an activation $a$ at layer $\ell$ is then $W_U\operatorname{norm}(J_\ell a)$. Since $J_\ell$ is averaged over many contexts, the readout reflects what the activation would generally lead the model to say, rather than what it leads to in one particular context. Gurnee et al. call the representations revealed by the J-lens \emph{verbalisable}, and the space they form the \emph{J-space}. They show that the J-space behaves like a global workspace. The model can report, hold, and reason with its contents, and pass them on to downstream computations. Although the J-space spans the entire vocabulary, only a few tokens are strongly active in it at any one time, yielding a sparse representation. In the models they study, these properties are most evident in the middle \emph{workspace} layers; readouts are noisy in the early layers and, in the final layers, mostly reflect the output the model is about to produce.

Like LIHS, J-lens readouts can be computed from the input alone, before any response is generated. They express the content of the model's workspace in vocabulary coordinates, with each feature corresponding to a token. We therefore investigate them as an input representation for pre-generation test adequacy.

\section{\name}
\label{sec:approach}

\name is a framework for pre-generation input adequacy: it supports two input representations, and two adequacy estimators. In addition to the Last-token Input Hidden States (LIHS) of Clotho, \name can also represent inputs using J-lens readouts~\cite{gurnee2026verbalizable}, which capture what the model is poised to verbalise. For estimators, \name supports Logistic Regression to predict pass or fail outcome of the output to be generated from the given input. Further, unlike Clotho, \name also learns from failing inputs, so that familiar failure patterns can be recognised, using dual GMM. 

Given a target model, a task-specific prompt template $T$, and a candidate pool $\mathcal P$, we extract a representation $h(x)=\phi(T(x))$ for each input $x$ without decoding a response. To label an input, we sample 10 responses and use a task-specific oracle to compute $u(x)$, the fraction judged correct. We assign $y(x)=\mathrm{pass}$ if a strict majority of the responses are correct and $y(x)=\mathrm{fail}$ otherwise. The labelled inputs form the \emph{reference set} $\mathcal R_t$ at acquisition round $t$, with passing and failing subsets $\mathcal R_{t,P}$ and $\mathcal R_{t,F}$. Once fitted on $\mathcal R_t$, the estimators score new inputs from $h(x)$ alone. Algorithm~\ref{alg:approach} constructs $\mathcal R_t$ adaptively, labelling new inputs and refitting the estimators in each round until the labelling budget is exhausted.

The remainder of this section describes input representations (Section~\ref{sec:representations}), adequacy estimators (Sections~\ref{sec:logistic} and~\ref{sec:adaptive-modelling}), and the adaptive construction of the reference set (Section~\ref{sec:reference-construction}).

\subsection{Input Representations}
\label{sec:representations}

We consider two representations extracted before response generation. \underline{L}ast-token \underline{I}nput \underline{H}idden \underline{S}tates (LIHS) use a model activation directly. \underline{J}acobian-lens \underline{R}eadout \underline{S}cores (JRS) is a vocabulary-aligned feature vector constructed from J-lens readouts. Either representation can be paired with either estimator.

\paragraph{LIHS}
Following Clotho~\cite{yoon2026clotho}, we extract the hidden state at the last token of the input prompt from the residual stream after the transformer layer located at two-thirds of the model's depth.

\paragraph{JRS}
Let $a_{\ell,p}(x)$ denote the residual-stream activation at layer $\ell$ and prompt position $p$. The J-lens maps intermediate activations through an averaged Jacobian $J_\ell$, followed by the model's final normalisation and unembedding~\cite{gurnee2026verbalizable}. We extract readouts at a fixed token position $p_{\mathrm J}$ in the chat-formatted input prompt, before any response tokens are generated. For each selected layer $\ell\in\mathcal L$, the resulting pre-softmax vocabulary scores are: $r_\ell(x) = W_U\operatorname{norm}\!\left( J_\ell a_{\ell,p_{\mathrm J}}(x) \right) \in\mathbb R^{|V|},$ where $W_U$ is the unembedding matrix and $V$ is the model vocabulary. The set of layers, $\mathcal L$, is chosen from a middle-to-late layer band $[\ell_{\mathrm{lo}},\ell_{\mathrm{hi}}]$, covering approximately $45\%$ to $90\%$ of each model's depth, following the workspace analysis of Gurnee et al.~\cite{gurnee2026verbalizable}. The exact layer bounds for each model are reported in the experimental setup.

To place readouts from different layers on a comparable scale, we standardize each $r_\ell(x)$ to zero mean and unit variance across its $|V|$ coordinates, obtaining $\widetilde r_\ell(x)$, and average over layers in $\mathcal{L}$: $\bar r(x)=\frac{1}{|\mathcal L|}\sum_{\ell\in\mathcal L}\widetilde r_\ell(x).$ Each coordinate $\bar r_v(x)$ is thus the standardised score of token $v\in V$, averaged over the selected layers. This pooling preserves the vocabulary alignment of the readouts but yields a $|V|$-dimensional vector, which is costly to model downstream. Because the informative variation in J-lens readouts tends to concentrate in a relatively small subset of vocabulary coordinates~\cite{gurnee2026verbalizable}, we retain only the set $V_K\subset V$ of $K=2000$ tokens with the largest variance $\operatorname{Var}_{x\in\mathcal P}[\bar r_v(x)]$ over the candidate pool, and use $[\bar r_v(x)]_{v\in V_K}\in\mathbb R^K$ as the representation. The selected set $V_K$ and its ordering are fixed and reused for all subsequent inputs.

In the remainder of this section, $h(x)$ denotes either the LIHS or JRS. Two adequacy estimators take the same $h(x)$ as input but differ in how they use the passing and failing references: the LR probe learns to separate two classes, whereas the dual GMM models the distribution of each class.

\subsection{Learning to Separate Pass from Fail using Logistic-Regression Estimator}
\label{sec:logistic}

The most direct way to score failure from the references is to learn a decision boundary between the two classes, i.e., pass and fail. At each round $t$, we fit an $\ell_1$-regularised logistic-regression probe to $\mathcal R_t$ on $h(x)$, and define its failure score as $s_{\mathrm L,t}(x)=\sigma\!\left(\mathbf w_t^{\top}h(x)+b_t\right),$ where $\sigma$ is the logistic function and $\mathbf w_t$ and $b_t$ are the fitted weights and bias. The $\ell_1$ penalty keeps the probe sparse to cater for the high dimensionality as well as the small size of the reference sets in the early rounds. %

\begin{algorithm}[t]
\small
\caption{Adaptive reference-set construction in \name, shown for the dual GMM}
\label{alg:approach}
\KwIn{candidate pool $\mathcal P$, prompt template $T$,
  representation $\phi$, initial reference size $N_0$,
  labelling budget $N$, batch size $B$, minimum references per class
  $m_{\min}$}
\KwOut{class models $\mathcal M_{t,P},\mathcal M_{t,F}$, or $\bot$ if either class has fewer than $m_{\min}$ references}
Construct and cache $h(x)=\phi(T(x))$ for all $x\in\mathcal P$\nllabel{ln:approach-features}\;
$\mathcal R_0\leftarrow\operatorname{RandomSample}(\mathcal P,N_0)$; label $\mathcal R_0$\nllabel{ln:approach-seed}\;
$\mathcal U_0\leftarrow\mathcal P\setminus\mathcal R_0$; $t\leftarrow0$\nllabel{ln:approach-init}\;
\While{\textnormal{true}}{
  $\mathcal R_{t,P}\leftarrow\{x\in\mathcal R_t:y(x)=\mathrm{pass}\}$;
  $\mathcal R_{t,F}\leftarrow\{x\in\mathcal R_t:y(x)=\mathrm{fail}\}$\nllabel{ln:approach-partition}\;
  \eIf{$\min(|\mathcal R_{t,P}|,|\mathcal R_{t,F}|)\geq m_{\min}$}{
    $(A_t,\theta_{t,P},\theta_{t,F})\leftarrow\operatorname{SelectConfigurations}(h(\mathcal P),\mathcal R_t)$\nllabel{ln:approach-select}\;
    \ForEach{$c\in\{P,F\}$}{
      $\pi_{t,c}\leftarrow|\mathcal R_{t,c}|/|\mathcal R_t|$\nllabel{ln:approach-priors}\;
      $\mathcal M_{t,c}\leftarrow\operatorname{FitDistribution}(h(\mathcal R_{t,c});A_t,\theta_{t,c},\pi_{t,c})$\nllabel{ln:approach-fit}\;
    }
  }{
    $\mathcal M_{t,P},\mathcal M_{t,F}\leftarrow\bot,\bot$\nllabel{ln:approach-unavailable}\;
  }
  \If{$|\mathcal R_t|\geq N$ \textnormal{or} $\mathcal U_t=\emptyset$}{
    \KwRet{$\mathcal M_{t,P},\mathcal M_{t,F}$}\nllabel{ln:approach-return}\;
  }
  $B_t\leftarrow\min(B,\,N-|\mathcal R_t|,\,|\mathcal U_t|)$\nllabel{ln:approach-budget}\;
  \eIf{$\mathcal M_{t,P}\neq\bot$ \textnormal{and} $\mathcal M_{t,F}\neq\bot$}{
    $B_{\mathrm{acq}}\leftarrow\lceil B_t/2\rceil$\nllabel{ln:approach-split}\;
    $C_{\mathrm{acq}}\leftarrow$ top-$B_{\mathrm{acq}}$ inputs $x\in\mathcal U_t$ by $H(s_{\mathrm G,t}(x))$\nllabel{ln:approach-acquire}\;
  }{
    $B_{\mathrm{acq}}\leftarrow0$; $C_{\mathrm{acq}}\leftarrow\emptyset$\nllabel{ln:approach-fallback}\;
  }
  $B_{\mathrm{div}}\leftarrow B_t-B_{\mathrm{acq}}$\nllabel{ln:approach-div-budget}\;
  $C_{\mathrm{div}}\leftarrow\operatorname{FarthestPointSample}(\mathcal U_t\setminus C_{\mathrm{acq}},\,\mathcal R_t\cup C_{\mathrm{acq}},\,B_{\mathrm{div}};h)$\nllabel{ln:approach-diversity}\;
  $C_t\leftarrow C_{\mathrm{acq}}\cup C_{\mathrm{div}}$; label $C_t$\nllabel{ln:approach-label}\;
  $\mathcal R_{t+1}\leftarrow\mathcal R_t\cup C_t$;
  $\mathcal U_{t+1}\leftarrow\mathcal U_t\setminus C_t$;
  $t\leftarrow t+1$\nllabel{ln:approach-update}\;
}
\end{algorithm}

\subsection{Modelling Passing and Failing Distributions using 
dual GMM Estimator}
\label{sec:adaptive-modelling}

The second estimator is dual Gaussian Mixture Models (dual GMMs), which is an extension of the GMM used by Clotho. The motivation for dual GMM is two-fold. First, the LR probe models only the boundary between the classes, pass and fail, but not how each class is distributed, providing no measure of how well the references support an input. Second, by modelling supports for both passing and failing inputs (hence the duality), \name aims to identify groups of similar failures.

The dual GMM projects the representations into a shared PCA space and fits a separate GMM to each outcome class. Together, the two class models yield three outputs: a posterior failure score comparable to that of the LR probe, a reference density that identifies inputs weakly supported by the references, and component memberships that group inputs within each class and provide a basis for investigating recurring failure patterns. We first define these outputs and then describe how the model configuration adapts as the reference set grows.

\subsubsection{Class models and derived scores.}
\label{sec:class-models}

At round $t$, both classes share a PCA transformation $A_t$, fitted on the candidate pool $\mathcal P$ and retaining $D_t$ dimensions, which maps each input to $z_t(x)=A_t h(x)\in\mathbb R^{D_t}$. Sharing the transformation places the passing and failing densities in the same coordinates; $A_t$ is stored with the fitted models and reused to score new inputs. In this space, we fit a Gaussian Mixture Model $p_{t,c}$ to each class $c\in\{P,F\}$ on $\mathcal R_{t,c}$, with a class-specific number of components and covariance structure selected as described below. With class priors $\pi_{t,c}=|\mathcal R_{t,c}|/|\mathcal R_t|$, the posterior failure score is:
\begin{equation}
s_{\mathrm G,t}(x)=
\frac{\pi_{t,F}\,p_{t,F}(z_t(x))}
{\pi_{t,P}\,p_{t,P}(z_t(x))+\pi_{t,F}\,p_{t,F}(z_t(x))}.
\label{eq:gmm-posterior}
\end{equation}
Higher scores indicate stronger evidence for failure relative to passing under the fitted models: we use this score to prioritise inputs before generation.

Modelling the class densities explicitly has a useful by-product with no counterpart in the LR probe: the normalizing term of Eq.~\eqref{eq:gmm-posterior} is itself informative. Its logarithm,
\begin{equation}
d_t(x)=\log\!\left[
\pi_{t,P}\,p_{t,P}(z_t(x))+\pi_{t,F}\,p_{t,F}(z_t(x))
\right],
\label{eq:reference-density}
\end{equation}
is the log marginal density of $z_t(x)$ under the fitted class models, which we call the \emph{reference density}. It measures how well the current references, pooled across both classes, support an input, regardless of which outcome the posterior favours. Computing the posterior already yields $d_t$, so the reference density comes at no additional cost; we compute both in log space for numerical stability. Pooling both classes distinguishes the reference density from the score by which Clotho ranks inputs: the density under a GMM fitted to passing references only. That score is low both for inputs that resemble failures and for inputs that resemble no reference, whereas the reference density is low only for the latter. Consequently, we can either prioritise inputs using the posterior failure score, or detect out-of-distribution inputs using the reference density.

Finally, each class mixture groups the inputs of its class into components: we assign an input to the component of $p_{t,c}$ with the highest posterior responsibility. Among failing references, these component memberships provide the basis for investigating recurring failure patterns.

\subsubsection{Adaptive configuration selection.}
\label{sec:adaptive-configuration}

As in Clotho~\cite{yoon2026clotho}, we adapt the configuration of the dual GMM as the reference set grows, since the appropriate model complexity depends on both the task and the number of available references. The configuration consists of the shared PCA dimensionality $D_t$ and, for each class $c$, the number of mixture components and the covariance structure, $\theta_{t,c}$. Searching over all combinations of these settings would require fitting many pairs of class models at every refit. Instead, each time the class models are refitted (line~\ref{ln:approach-select} of Algorithm~\ref{alg:approach}), we update the configuration in two steps, each a local search around the previous configuration: we first select the dimensionality with the component counts held fixed, and then reselect the component counts for the chosen dimensionality.

First, we select the dimensionality. The first fit uses 30 dimensions, since no component counts are yet available for evaluating alternatives; at each subsequent refit, we compare the previous dimensionality with $\pm$10 dimensions. Using 3-fold cross-validation on $\mathcal R_t$, we fit the class models to the training references, compute the passing log odds of each held-out reference $x$, $\log[\pi_P\,p_P(z(x))/\pi_F\,p_F(z(x))]$, and compute their Spearman correlation to the observed success rates $u(x)$. We keep the dimensionality with the highest mean correlation across folds.

Second, we reselect the number of components of each class using the newly chosen dimensionality. Ideally, this number matches the number of clusters that the inputs of the class form, but that number is not only unknown in advance but also depends on the dimensionality as well as the reference set. For each class, the number of components we choose is the one that maximises the 3-fold cross-validated log likelihood of the class reference set, i.e., one that fits the existing references the best. Note that, since log likelihoods cannot be compared across different dimensions, the number of components is chosen second. It is also capped at $\min(10,\lfloor|\mathcal R_{t,c}|/10\rfloor)$, so that each component is fitted to at least ten references on average. The first fit considers every count up to this cap; subsequent refits consider only the previous count and its $\pm1$ neighbouring values. We use full covariance matrices by default and fall back to diagonal ones whenever a class has too few references to estimate them, i.e., fewer references than the free parameters of the full-covariance.

\subsection{Adaptive Reference-Set Construction}
\label{sec:reference-construction}

Algorithm~\ref{alg:approach} constructs the reference set iteratively, refitting the estimator as new labels are acquired. We present it for the dual GMM and describe the LR variant at the end of this section. Given the candidate pool $\mathcal P$ and a labelling budget $N$, the algorithm first constructs and caches the representations of all candidates (line~\ref{ln:approach-features}). It then labels $N_0$ randomly selected inputs as the initial references and initialises the unlabelled pool (lines~\ref{ln:approach-seed}--\ref{ln:approach-init}).  At each round $t$, it partitions the references by outcome (line~\ref{ln:approach-partition}). Once both classes contain at least $m_{\min}$ references, it selects the PCA transformation and the class-specific mixture configurations $\theta_{t,P}$ and $\theta_{t,F}$ as described in Section~\ref{sec:adaptive-modelling} (line~\ref{ln:approach-select}), estimates the class priors, and fits the class models $\mathcal M_{t,P}$ and $\mathcal M_{t,F}$ (lines~\ref{ln:approach-priors}--\ref{ln:approach-fit}). Otherwise, no fitted scorer is available (line~\ref{ln:approach-unavailable}). The stopping check follows this fit: if the budget is exhausted or the unlabelled pool is empty, the algorithm returns the scorer fitted on the current reference set, or $\bot$ if either class still lacks $m_{\min}$ references (line~\ref{ln:approach-return}). If acquisition continues, the next batch size $B_t$ is limited by the remaining budget and pool size (line~\ref{ln:approach-budget}). When the scorer is available, $\lceil B_t/2\rceil$ inputs are selected by the model-based acquisition criterion \textsc{Margin} (lines~\ref{ln:approach-split}--\ref{ln:approach-acquire}), which ranks unlabelled inputs by the binary entropy $H(s_{\mathrm G,t}(x))$ of their posterior failure scores, favouring inputs near the fitted pass/fail boundary.  Diversity sampling fills the remaining positions by repeatedly choosing the input with the greatest minimum Euclidean distance in $h$-space from the references and inputs already selected for this batch (lines~\ref{ln:approach-div-budget}--\ref{ln:approach-diversity}). If the scorer is unavailable, diversity sampling selects the entire batch (line~\ref{ln:approach-fallback}). The batch is then labelled, added to the references, and removed from the unlabelled pool before the next round (lines~\ref{ln:approach-label}--\ref{ln:approach-update}). The LR variant uses the same acquisition loop. Lines~\ref{ln:approach-select}--\ref{ln:approach-fit} instead fit the probe defined in Section~\ref{sec:logistic} to $\mathcal R_t$, and \textsc{Margin} uses $s_{\mathrm L,t}$ in place of $s_{\mathrm G,t}$.

\section{Experimental Setup}
\label{sec:experimental_setup}

This section details the experimental setup used to evaluate \name.

\subsection{Research Questions}
We organise the study around the following research questions.

\subsubsection{RQ1. How effectively does \name predict failing inputs and prioritise them for testing?} 
\label{sec:rq1-setup}

We evaluate each method on held-out inputs drawn from the same task. This assesses whether pre-generation scores separate failure-inducing inputs from passing ones, and whether ranking inputs by predicted failure risks enables earlier failure discovery under limited execution budgets.

\paragraph{Compared Methods.} 
We compare \name with the pre-generation baseline Clotho~\cite{yoon2026clotho} and three post-generation adequacy metrics: \emph{Token Probability} (\tokp) and \emph{Token Entropy} (\toke)~\cite{Manakul2023aa}, and \emph{Last Output-token Hidden State Variance} (\lohsv). We compare \name to all the above for evaluation of failure prediction, and only to pre-generation methods as well as random for evaluation of prioritisation. We study the 2 by 2 combinations of representations (LIHS and JRS) and estimators (LR and dual GMM) to separate their contributions.

\paragraph{Evaluation Metrics.}
We report ROC-AUC for failure prediction study, which measures how well risk scores rank inputs labelled \textit{fail} above those labelled \textit{pass}. For prioritisation, we report Failure@$N$, which counts failed inputs among the $N\in\{100,300,500\}$ highest-risk candidates. For both, we report means over ten seeds, and provide details in our replication package~\cite{replication}.

\subsubsection{RQ2. How reliably does \name identify inputs that fall outside its task-specific distribution?}
\label{sec:rq2-setup}

Before acting on a learned failure score, a tester may ask: \emph{Can I trust this score for this input?} For this, we evaluate whether studied estimators can reliably distinguish inputs within the task distribution from those out of distribution, based on the small set of labelled reference inputs. 

\paragraph{Evaluation Setup.} 
For each target task, we partition inputs into three disjoint sets: reference, calibration, and evaluation. The references fit the estimators; the remaining target-task inputs, both passing and failing, are divided roughly evenly between calibration and evaluation. We sample 500 out-of-distribution inputs from four randomly chosen remaining tasks, and inject their input into the target task's prompt template to extract the same representation.

\paragraph{Compared Methods.}
For each seed, we fit a dual GMM and an $\ell_1$-regularised LR probe on the same labelled LIHS references, using each estimator's preprocessing procedure. We compare the dual GMM marginal reference log-density $d_t(x)$ (Eq.~\eqref{eq:reference-density}), and the LR pass probability $P(\text{pass}\mid x)$.\footnote{The intuition for LR scores is that an input confidently predicted to pass must be in-distribution for the target task.}

\paragraph{Evaluation Metrics.}
For each target, seed, and score, we set a threshold $\tau$ to retain $95\%$ of the in-distribution calibration inputs, then report the False Positive Rate (FPR)@95 when we apply $\tau$ to the held-out target task inputs as well as OOD inputs. We also compute ROC-AUC for each OOD source against the held-out target inputs, treating the latter as positive. FPR@95 is the lower the better, ROC-AUC the higher the better. We report means and standard deviations over ten seeds.

\subsubsection{RQ3. Can \name help characterise LLM failures?}
\label{sec:rq3-setup}
As JRS is tied to specific tokens, we can represent observed failure patterns as tokens. This is more straightforward with failure components of dual GMM: we can pool all inputs belonging to each failure component, and apply J-lens to read the tokens. LR, on the other hand, simply fits a hyperplane in the JRS space which best separates passing and failing input representations. Consequently, we can read out the tokens that have the highest influence on the separation from the normalised weights of the fitted LR models. We qualitatively analyse whether such readouts can characterise different groups of failures taken from \topic and \mbpp tasks using Llama-3.1-8B.

\paragraph{Failure Patterns.} 
On \topic, we examine the failing-class GMM components through two complementary analyses. First, we assess whether they show distinctive failure patterns, characterised by each input's ground truth topic and the predicted topic. Second, we examine whether the tokens we read out of the failing input representation reveal semantic characteristics of the failure pattern.
To evaluate whether components group failures with similar observed patterns, we measure associations between component membership and ground-truth topics, predicted topics, and true--predicted topic pairs using Cram\'er's $V$~\cite{cramer1946mathematical}. %
We report means and standard deviations over ten random seeds, and illustrate the patterns using one run (seed~0).

\paragraph{Semantic Characterisation.}
For each input $x$ and token $v\in V_K$, we average the token's J-lens readout across the workspace bands to obtain $\bar r_v(x)$, i.e., JRS. We then rank tokens by the difference between the mean of $\bar r_v(x)$ over $F_j$ and its mean over the entire input set, to highlight tokens elevated in $F_j$. After merging case and punctuation variants, we inspect the top 20 tokens of each component, identify the concept they express, and examine how its prominence in the model's representation relates to the component's failures. For \topic, at the task level, we also rank tokens by their weights in the LR, separately for positive weights (towards failure) and negative weights (towards passing), and inspect the top 20 of each after the same merging. Finally, to show how \name can assist analysis of individual cases, we report top tokens from failing inputs randomly sampled from held-out sets, i.e., those not labelled during GMM fitting. We do this on both \topic and \mbpp. For \mbpp, we also compare the readouts with the generated programs and reference solutions.

\subsubsection{RQ4. How effectively can \name scores obtained from smaller open-weight models prioritize failures in larger or proprietary target models?}
\label{sec:rq4-setup}

\name requires hidden states, which are not available from proprietary models and expensive to extract from larger open-weight models. Clotho showed that pass-only scores from smaller open-weight models transfer to proprietary models~\cite{yoon2026clotho}; we ask whether better modelling of failures in smaller models can improve these transfers. To test this, we fit estimators using source representations from smaller models, and use target outcomes to evaluate the orderings obtained using the estimator scores (see Section~\ref{sec:models} for model pairs).

\paragraph{Compared Methods.} We compare all RQ1 methods with Clotho and random ordering.

\paragraph{Evaluation Metrics.}
Following Clotho, we use a fixed random subset of 2,000 inputs per task, or all non-training inputs for the smaller \mbpp and \live datasets. The inputs are shared across sources, targets, methods, and seeds. Each estimator prioritises them from the most to the least likely to fail. We report normalised \fan: the number of failing inputs among the first $N$ = 25\% of the ordering, divided by $\min(F, N)$, where $F$ is the number of failing inputs in the evaluation set. A perfect prioritisation scores \fan of 1.0, whereas random prioritisation would on average score 0.25. Results are averaged over ten seeds.

\subsubsection{RQ5. What computational cost does \name incur?}
\label{sec:rq5-setup}

We measure computational cost on three randomly chosen tasks (\model, \pos, and \topic) at a fixed seed. Because representations are cached and reused across fitting runs, we time extraction and fitting separately. All timings are taken on one machine with a single NVIDIA GeForce RTX 3090 GPU at batch size~1. 

\paragraph{Compared Methods.} 

We compare computational costs of all methods studied in RQ1. Each estimator uses the same configuration and active-learning loop as in RQ1.

\paragraph{Evaluation Metrics.}

For LIHS and JRS, we time the extraction of the same 25 inputs per dataset, three times each, and report the mean and standard deviation of per-input wall-clock time, which for JRS includes the readout operations. For fitting, we compare Clotho's passing-class GMM with full covariance on LIHS~\cite{yoon2026clotho}, the dual GMM, and the LR probe, fitting the latter two on both LIHS and J-lens, and report wall-clock times.

\subsection{Experimental Details}
\label{sec:details}
 The experimental details for our study is provided.

\begin{table}[t]
\centering
\caption{Target tasks, datasets, and pass rates ($p = 0.5$)
of the studied open-weight LLMs.}
\label{tab:tasks_pass_rates}
\resizebox{0.8\columnwidth}{!}{%
\begin{tabular}{@{}lrlrrr@{}}
\toprule
\textbf{Task} & \textbf{Inputs} & \textbf{Description}
& \textbf{Llama 3.1 8B}
& \textbf{Gemma 2 9B} & \textbf{Qwen 3 8B} \\
\midrule
\texttt{ODD-ADD}
& 6,000
& Odd-number summation~\cite{yoon2026clotho}
& 0.60 & 0.60 & 0.69 \\

\texttt{GH-TYPO}
& 10,000
& Github Typo correction~\cite{hagiwara2019github}
& 0.51 & 0.64 & 0.58 \\

\json
& 6,563
& Repair invalid JSON ~\cite{github:jsonrepair}
& 0.75 & 0.83 & 0.94 \\

\texttt{MODEL-EX}
& 9,810
& Extract ML model names~\cite{huggingface:arxiv_abstracts_dataset}
& 0.55 & 0.67 & 0.75 \\

\texttt{POS-TAG}
& 15,359
& Detect Part-of-speech tags~\cite{silveira14gold}
& 0.77 & 0.81 & 0.79 \\

\spell
& 10,000
& Fix misspelt words~\cite{fellbaum1998wordnet,mitton1980birkbeck}
& 0.82 & 0.88 & 0.86 \\

\texttt{SYN-BUG}
& 20,518
& Find code lines with syntactic bugs~\cite{lee2024bug}
& 0.36 & 0.42 & 0.47 \\

\topic
& 7,600
& Classify topics of news articles~\cite{Zhang2015CharacterlevelCN}
& 0.78 & 0.82 & 0.77 \\

\midrule
\mbpp
& 974
& Python synthesis (MBPP~\cite{austin2021program})
& 0.63 & 0.64 & 0.71 \\

\live
& 1,055
& Python synthesis (LiveCodeBench~\cite{jain2025livecodebench})
& 0.20 & 0.26 & 0.50 \\
\bottomrule
\end{tabular}%
}

\end{table}

\subsubsection{Target Tasks and Datasets}
\label{sec:tasks}

We evaluate \name on the eight benchmark tasks used in Clotho~\cite{yoon2026clotho}, as well as two python program synthesis benchmarks: Mostly Basic Python Problems (MBPP)~\cite{austin2021program} and LiveCodeBench~\cite{jain2025livecodebench}. Table~\ref{tab:tasks_pass_rates} summarizes the tasks and their corresponding datasets.

\subsubsection{Subject Models and Transfer Targets}
\label{sec:models} 
We evaluate \name using Llama 3.1 8B~\cite{llama3modelcard}, Gemma 2 9B~\cite{gemma_2024}, and Qwen 3 8B~\cite{qwen3technicalreport}, for which pretrained J-lens models are publicly available~\cite{neuronpedia_jacobian_lens}. The last three columns of table~\ref{tab:tasks_pass_rates} reports the fraction of inputs each model passes on each task. For RQ4, we additionally evaluate larger models from the same families -- Llama 3.3 70B~\cite{llama3modelcard}, Gemma 3 27B~\cite{gemma_2024}, and Qwen 3 32B~\cite{qwen3technicalreport} -- to investigate the effect of model scale. We further include three proprietary models, GPT-4o mini~\cite{openai_gpt4omini_docs}, Gemini 3.5 Flash Lite~\cite{google_gemini_35_flash_lite_docs}, and Claude Haiku 4.5~\cite{anthropic2025haiku45}.

\subsubsection{Evaluation Protocol and Implementation Details}
\label{sec:eval-protocol}
The reference set is initialised with ten inputs, sampled uniformly at random from a pool of unlabelled inputs and then labelled, and is subsequently expanded in increments of ten until a total labelling budget of $N = 500$ inputs is reached.

We implement the logistic regression probe and the density estimators using \texttt{LogisticRegression} from \texttt{sklearn.linear\_model} and \texttt{GaussianMixture} from \texttt{sklearn.mixture}, respectively~\cite{Pedregosa2011tk}. For logistic regression, we use the \texttt{liblinear} solver. We tune the inverse regularisation strength $C$ by cross-validation, choosing the value with the highest AUROC among ten values between $10^{-2}$ and $10$. When there are too few references for cross-validation, we set $C = 1$.
For density estimation, separate GMMs are fitted to the passing and failing references in a shared PCA space, and their configuration is selected adaptively as described in Section~\ref{sec:adaptive-configuration}. We restrict the PCA dimensionality to the range of 10 to 50. When cross-validation scores are unavailable for a class, we select its number of components using the Bayesian information criterion (BIC) instead. Expectation--maximisation terminates when the change in the average log-likelihood lower bound falls below $10^{-3}$ or after 200 iterations, whichever occurs first.

To construct JRS, we extract representations at the model-specific assistant-header position immediately preceding response generation, and average per-layer-standardised Jacobian-lens readouts over a mid-to-late workspace band: layers 14–28 of 32 for Llama-3.1-8B, 18–37 of 42 for Gemma-2-9B, and 16–32 of 36 for Qwen3-8B (0-indexed, inclusive).

\section{Results}
\label{sec:results}

This section presents the results of our experiments.

\begin{table*}[t]
\centering
\caption{ROC-AUC for pre-generation pass/fail prediction (\textbf{Best}, \underline{second best})}
\label{tab:rq1-rocauc}
\resizebox{\textwidth}{!}{%
\begin{tabular}{l|rrrrr|rrrrr|rrrrr}
\toprule
 & \multicolumn{5}{c|}{Gemma-2-9B} & \multicolumn{5}{c|}{Llama-3.1-8B} & \multicolumn{5}{c}{Qwen3-8B} \\
\cmidrule(lr){2-6} \cmidrule(lr){7-11} \cmidrule(lr){12-16}
Task ID & Clotho & LIHS$_{\mathrm{DG}}$ & LIHS$_{\mathrm{LR}}$ & JRS$_{\mathrm{DG}}$ & JRS$_{\mathrm{LR}}$ & Clotho & LIHS$_{\mathrm{DG}}$ & LIHS$_{\mathrm{LR}}$ & JRS$_{\mathrm{DG}}$ & JRS$_{\mathrm{LR}}$ & Clotho & LIHS$_{\mathrm{DG}}$ & LIHS$_{\mathrm{LR}}$ & JRS$_{\mathrm{DG}}$ & JRS$_{\mathrm{LR}}$ \\
\midrule
\odd & 0.682 & 0.799 & \textbf{0.838} & 0.809 & \underline{0.836} & 0.811 & 0.880 & \textbf{0.901} & 0.858 & \underline{0.896} & 0.726 & 0.792 & \underline{0.828} & 0.818 & \textbf{0.837} \\
\gh & 0.676 & 0.811 & \underline{0.838} & 0.790 & \textbf{0.851} & 0.723 & 0.807 & \underline{0.848} & 0.801 & \textbf{0.848} & 0.621 & 0.723 & \underline{0.813} & 0.779 & \textbf{0.839} \\
\json & 0.726 & 0.818 & \textbf{0.866} & 0.752 & \underline{0.859} & 0.721 & 0.758 & \textbf{0.823} & 0.764 & \underline{0.812} & 0.595 & 0.709 & \underline{0.800} & 0.751 & \textbf{0.828} \\
\model & 0.624 & 0.848 & \textbf{0.890} & 0.846 & \underline{0.873} & 0.673 & 0.820 & \underline{0.858} & 0.782 & \textbf{0.859} & 0.471 & 0.720 & \underline{0.773} & 0.723 & \textbf{0.785} \\
\pos & 0.631 & 0.628 & 0.708 & \underline{0.712} & \textbf{0.740} & 0.561 & 0.622 & \underline{0.697} & 0.683 & \textbf{0.731} & 0.596 & 0.674 & \underline{0.745} & 0.668 & \textbf{0.752} \\
\spell & 0.855 & 0.886 & \underline{0.905} & 0.894 & \textbf{0.919} & 0.747 & 0.871 & \textbf{0.894} & 0.839 & \underline{0.890} & 0.693 & 0.838 & \textbf{0.879} & 0.816 & \underline{0.878} \\
\syn & 0.888 & 0.936 & \textbf{0.954} & 0.887 & \underline{0.943} & 0.718 & 0.849 & 0.928 & \underline{0.931} & \textbf{0.950} & 0.744 & 0.868 & 0.895 & \underline{0.918} & \textbf{0.923} \\
\topic & 0.607 & 0.807 & 0.816 & \textbf{0.864} & \underline{0.864} & 0.614 & 0.803 & 0.808 & \textbf{0.854} & \underline{0.827} & 0.563 & 0.717 & 0.781 & \textbf{0.847} & \underline{0.830} \\
\mbpp & 0.639 & 0.684 & \textbf{0.798} & 0.689 & \underline{0.759} & 0.643 & 0.733 & \textbf{0.843} & 0.750 & \underline{0.828} & 0.593 & 0.733 & \textbf{0.812} & 0.708 & \underline{0.807} \\
\live & 0.623 & \underline{0.749} & \textbf{0.831} & 0.714 & 0.695 & 0.786 & 0.785 & \textbf{0.903} & 0.708 & \underline{0.841} & 0.707 & \underline{0.846} & \textbf{0.891} & 0.774 & 0.798 \\
\midrule
Average & 0.695 & 0.797 & \textbf{0.844} & 0.796 & \underline{0.834} & 0.700 & 0.793 & \textbf{0.850} & 0.797 & \underline{0.848} & 0.631 & 0.762 & \underline{0.822} & 0.780 & \textbf{0.828} \\
\bottomrule
\end{tabular}%
}
\end{table*}

\subsection{RQ1: In-distribution Failure Prediction and Test Prioritisation}
\label{sec:rq1}

\subsubsection{RQ1-1: Failure Prediction.}
\label{sec:rq1-1}
Table~\ref{tab:rq1-rocauc} reports ROC-AUC with 500 labelled references. All four configurations of \name separate failing from passing inputs better than Clotho: their mean ROC-AUC is 0.784 for LIHS$_{\mathrm{DG}}$, 0.791 for JRS$_{\mathrm{DG}}$, 0.839 for LIHS$_{\mathrm{LR}}$, and 0.837 for JRS$_{\mathrm{LR}}$, compared with 0.675 for Clotho. The improvement holds on every task: averaged over LLMs, each configuration exceeds Clotho on all ten tasks, and both LR configurations do so in all 30 task--LLM combinations. Part of this gain may reflect label usage: Clotho models only the passing-input distribution and discards the failing examples during label acquisition, whereas \name learns from both.

To assess estimator and representation effects, we use four paired contrasts: LR versus dual GMM within each representation, and JRS versus LIHS within each estimator. We treat tasks as the unit of inference, averaging each contrast over seeds and the three LLMs within a task. For the ten resulting differences, we report a Hodges--Lehmann (HL) estimate and 95\% confidence interval, test against zero with a two-sided exact Wilcoxon signed-rank test, and apply Holm's correction to the four $p$-values for each metric. Estimator choice shows the clearer pattern. LR gains about $0.05$ ROC-AUC over dual GMM with either representation (Holm-adjusted $p\leq 0.012$), leading on every task with LIHS and all but \topic with JRS. The JRS--LIHS contrasts are near zero (ROC-AUC +$0.003$) under both estimators, with confidence intervals spanning zero; we find no clear overall representation advantage. 
On the two program-synthesis tasks, \mbpp and \live, LIHS$_{\mathrm{LR}}$ achieves mean ROC-AUCs of 0.82 and 0.88, versus 0.80 and 0.78 for JRS$_{\mathrm{LR}}$ and 0.63 and 0.71 for Clotho. JRS$_{\mathrm{LR}}$ instead outperforms LIHS$_{\mathrm{LR}}$ on \topic and \pos, showing that the preferred representation varies by task.

Table~\ref{tab:rq1-postgen} compares LIHS$_{\mathrm{LR}}$ with three post-generation signals for pass/fail prediction. Across the 30 task--model combinations, LIHS$_{\mathrm{LR}}$ achieves the highest mean ROC-AUC of 0.839, ahead of LOHS$_{\mathrm{var}}$ at 0.750, TOK$_{\mathrm{ent}}$ at 0.680, and TOK$_{\mathrm{prob}}$ at 0.616. It outperforms every post-generation signal in 22 of the 30 combinations. Note that the methods have different requirements: the post-generation signals use generated outputs but need no reference labels, whereas LIHS$_{\mathrm{LR}}$ is trained on $N=500$ labelled examples and can score inputs before generation.

\begin{table*}[t]
\centering
\caption{ROC-AUC for post-generation pass/fail prediction, with LIHS$_{\mathrm{LR}}$ as the pre-generation reference.}
\label{tab:rq1-postgen}
\resizebox{\textwidth}{!}{%
\begin{tabular}{l|rrrr|rrrr|rrrr}
\toprule

Task ID & \multicolumn{4}{c}{Gemma-2-9B} & \multicolumn{4}{c}{Llama-3.1-8B} & \multicolumn{4}{c}{Qwen3-8B} \\
\cmidrule(lr){2-5} \cmidrule(lr){6-9} \cmidrule(lr){10-13}
 & LOHS$_{\mathrm{var}}$ & TOK$_{\mathrm{prob}}$ & TOK$_{\mathrm{ent}}$ & LIHS$_{\mathrm{LR}}$ & LOHS$_{\mathrm{var}}$ & TOK$_{\mathrm{prob}}$ & TOK$_{\mathrm{ent}}$ & LIHS$_{\mathrm{LR}}$ & LOHS$_{\mathrm{var}}$ & TOK$_{\mathrm{prob}}$ & TOK$_{\mathrm{ent}}$ & LIHS$_{\mathrm{LR}}$ \\
\midrule
\odd & 0.742 & 0.434 & \underline{0.755} & \textbf{0.838} & 0.705 & \underline{0.764} & 0.714 & \textbf{0.901} & 0.723 & 0.825 & \textbf{0.831} & \underline{0.828} \\
\gh & \textbf{0.868} & 0.657 & 0.800 & \underline{0.838} & \textbf{0.868} & 0.710 & 0.755 & \underline{0.848} & \textbf{0.870} & 0.829 & \underline{0.857} & 0.813 \\
\json & \underline{0.736} & 0.545 & 0.604 & \textbf{0.866} & \textbf{0.827} & 0.628 & 0.645 & \underline{0.823} & \underline{0.617} & 0.531 & 0.539 & \textbf{0.800} \\
\model & \underline{0.808} & 0.391 & 0.581 & \textbf{0.890} & \underline{0.846} & 0.599 & 0.611 & \textbf{0.858} & \textbf{0.776} & 0.547 & 0.558 & \underline{0.773} \\
\pos & \textbf{0.772} & 0.526 & 0.627 & \underline{0.708} & \textbf{0.803} & 0.628 & 0.651 & \underline{0.697} & \underline{0.711} & 0.596 & 0.623 & \textbf{0.745} \\
\spell & \underline{0.896} & 0.705 & 0.831 & \textbf{0.905} & \underline{0.884} & 0.755 & 0.785 & \textbf{0.894} & \underline{0.845} & 0.714 & 0.755 & \textbf{0.879} \\
\syn & 0.525 & 0.522 & \underline{0.569} & \textbf{0.954} & 0.689 & 0.670 & \underline{0.697} & \textbf{0.928} & 0.577 & 0.573 & \underline{0.612} & \textbf{0.895} \\
\topic & \underline{0.631} & 0.535 & 0.578 & \textbf{0.816} & \underline{0.762} & 0.584 & 0.589 & \textbf{0.808} & \underline{0.684} & 0.569 & 0.583 & \textbf{0.781} \\
\mbpp & \underline{0.636} & 0.417 & 0.608 & \textbf{0.798} & 0.719 & 0.715 & \underline{0.732} & \textbf{0.843} & \underline{0.713} & 0.675 & 0.687 & \textbf{0.812} \\
\live & \underline{0.806} & 0.398 & 0.765 & \textbf{0.831} & 0.613 & 0.616 & \underline{0.640} & \textbf{0.903} & \underline{0.842} & 0.809 & 0.818 & \textbf{0.891} \\
\midrule
Average & \underline{0.742} & 0.513 & 0.672 & \textbf{0.844} & \underline{0.772} & 0.667 & 0.682 & \textbf{0.850} & \underline{0.736} & 0.667 & 0.686 & \textbf{0.822} \\
\bottomrule
\end{tabular}%
}
\end{table*}

\subsubsection{RQ1-2: Test Prioritisation}
\label{sec:rq1-2}

To assess whether these discrimination gains translate into earlier failure discovery, Table~\ref{tab:rq1-failure-at-n} reports failure@$N$ for each task and LLM. Across the 84 task--LLM--$N$ combinations (\mbpp and \live excluded for $N=500$ due to small dataset size), each configuration finds more failure-inducing inputs than Clotho in 63 to 76 of them (LIHS$_{\mathrm{LR}}$: 76) and more than random ordering in at least 83. These results demonstrate the practical benefit of the learned scores: testers can discover more failures on average within the same execution budget.

\begin{rqbox}[Answer to RQ1]
\name supports pre-generation failure prediction and earlier failure discovery on held-out inputs from the same task distribution. Across both representations and estimators, the learned scores outperform Clotho on average in difficulty ranking, pass/fail discrimination, and failure discovery within a fixed execution budget. LR provides the most consistent gains, while representation advantages depend on the task and model.
\end{rqbox}

\begin{table*}[ht]
\centering
\caption{Mean Failure@$N$ ($N \in \{100, 300, 500\}$) for pre-generation methods and baselines.}
\label{tab:rq1-failure-at-n}
\resizebox{\textwidth}{!}{%
\begin{tabular}{ll|rrrrrr|rrrrrr|rrrrrr}
\toprule
\multicolumn{2}{c}{} & \multicolumn{6}{c}{Gemma-2-9B} & \multicolumn{6}{c}{Llama-3.1-8B} & \multicolumn{6}{c}{Qwen3-8B} \\
\cmidrule(lr){3-8} \cmidrule(lr){9-14} \cmidrule(lr){15-20}
Task ID & $N$ & Clotho & LIHS$_{\mathrm{DG}}$ & LIHS$_{\mathrm{LR}}$ & JRS$_{\mathrm{DG}}$ & JRS$_{\mathrm{LR}}$ & Rand. & Clotho & LIHS$_{\mathrm{DG}}$ & LIHS$_{\mathrm{LR}}$ & JRS$_{\mathrm{DG}}$ & JRS$_{\mathrm{LR}}$ & Rand. & Clotho & LIHS$_{\mathrm{DG}}$ & LIHS$_{\mathrm{LR}}$ & JRS$_{\mathrm{DG}}$ & JRS$_{\mathrm{LR}}$ & Rand. \\
\midrule
\multirow{3}{*}{\odd} & 100 & \textbf{100.0} & \textbf{100.0} & \textbf{100.0} & \textbf{100.0} & \textbf{100.0} & 38.4 & 99.2 & 93.7 & 99.0 & 90.4 & \textbf{99.3} & 44.3 & 83.7 & 94.5 & 93.4 & 73.1 & \textbf{95.4} & 36.8 \\
 & 300 & 299.9 & 299.9 & 299.7 & \textbf{300.0} & 298.2 & 115.5 & 292.9 & 278.5 & 296.0 & 272.0 & \textbf{296.1} & 132.2 & 228.3 & 263.4 & 265.2 & 234.0 & \textbf{270.5} & 99.4 \\
 & 500 & 463.9 & 472.1 & 479.0 & \textbf{484.8} & 481.6 & 193.3 & 476.5 & 461.9 & 487.9 & 446.8 & \textbf{488.4} & 221.7 & 360.2 & 413.6 & 424.9 & 394.7 & \textbf{431.9} & 165.2 \\
\midrule
\multirow{3}{*}{\gh} & 100 & 83.3 & 89.9 & 89.9 & 90.2 & \textbf{94.4} & 35.1 & 91.9 & 96.1 & \textbf{98.3} & 91.1 & 95.1 & 52.3 & 73.2 & 86.2 & 94.1 & 87.0 & \textbf{96.1} & 42.3 \\
 & 300 & 223.6 & 265.5 & 267.4 & 268.5 & \textbf{278.5} & 106.0 & 272.4 & 291.7 & \textbf{295.1} & 279.0 & 288.3 & 151.8 & 211.4 & 246.1 & 273.2 & 260.1 & \textbf{280.6} & 128.5 \\
 & 500 & 358.5 & 439.5 & 441.0 & 441.2 & \textbf{455.7} & 182.2 & 448.3 & 484.4 & \textbf{489.7} & 465.3 & 480.3 & 254.9 & 337.9 & 404.1 & 447.3 & 427.9 & \textbf{462.9} & 210.8 \\
\midrule
\multirow{3}{*}{\json} & 100 & 86.1 & \textbf{90.2} & 84.8 & 88.8 & 82.2 & 20.0 & 94.2 & 90.1 & 93.6 & 94.1 & \textbf{94.5} & 32.3 & 13.2 & 39.6 & 60.1 & \textbf{70.3} & 60.6 & 6.2 \\
 & 300 & 237.3 & 262.1 & 244.3 & \textbf{263.0} & 235.2 & 63.3 & 249.6 & 253.6 & 258.8 & 258.1 & \textbf{265.0} & 100.4 & 32.1 & 84.3 & 137.3 & 140.5 & \textbf{142.7} & 19.5 \\
 & 500 & 342.3 & 393.3 & 386.6 & \textbf{424.6} & 383.8 & 104.5 & 367.3 & 379.8 & \textbf{408.2} & 386.8 & 407.9 & 165.5 & 51.1 & 117.2 & 178.6 & 171.9 & \textbf{193.5} & 35.7 \\
\midrule
\multirow{3}{*}{\model} & 100 & 59.5 & 85.7 & \textbf{96.8} & 87.5 & 94.5 & 34.0 & 81.0 & 86.8 & \textbf{95.1} & 82.3 & 91.1 & 48.2 & 23.1 & 50.3 & 73.3 & 62.2 & \textbf{79.3} & 26.1 \\
 & 300 & 165.6 & 248.7 & \textbf{284.5} & 264.8 & 278.0 & 100.3 & 226.6 & 251.2 & \textbf{279.3} & 245.7 & 268.5 & 143.4 & 69.7 & 151.5 & 204.2 & 189.0 & \textbf{223.3} & 79.5 \\
 & 500 & 268.3 & 405.7 & \textbf{463.0} & 430.9 & 455.4 & 167.2 & 367.0 & 415.2 & \textbf{458.8} & 407.2 & 443.7 & 235.7 & 114.3 & 251.2 & 328.2 & 305.7 & \textbf{357.6} & 134.4 \\
\midrule
\multirow{3}{*}{\pos} & 100 & 84.9 & 84.3 & 80.3 & \textbf{87.7} & 85.5 & 20.5 & 76.3 & 74.5 & \textbf{81.4} & 63.7 & 74.7 & 26.7 & 81.6 & 84.9 & \textbf{91.2} & 77.3 & 78.0 & 22.1 \\
 & 300 & 220.2 & 240.4 & 208.3 & \textbf{255.8} & 234.6 & 60.5 & 175.3 & 189.8 & \textbf{219.3} & 186.7 & 211.7 & 76.8 & 189.8 & 205.9 & \textbf{237.1} & 217.4 & 221.3 & 67.8 \\
 & 500 & 319.7 & 341.0 & 326.2 & \textbf{386.1} & 368.0 & 102.7 & 253.6 & 293.1 & \textbf{342.7} & 314.7 & 335.3 & 128.2 & 278.6 & 311.3 & \textbf{354.1} & 344.7 & 350.2 & 114.7 \\
\midrule
\multirow{3}{*}{\spell} & 100 & 85.7 & 85.0 & \textbf{88.9} & 83.2 & 84.7 & 12.0 & 76.5 & 82.4 & \textbf{90.1} & 79.6 & 86.2 & 19.1 & 75.7 & \textbf{81.3} & 81.1 & 78.5 & 79.8 & 13.5 \\
 & 300 & 234.1 & 232.0 & \textbf{244.5} & 229.7 & 242.1 & 34.4 & 214.4 & 243.3 & \textbf{255.4} & 230.7 & 251.8 & 57.5 & 194.7 & 228.5 & \textbf{234.8} & 223.8 & 227.9 & 41.9 \\
 & 500 & 356.8 & 355.8 & 379.8 & 356.5 & \textbf{379.9} & 57.5 & 329.2 & 392.0 & \textbf{414.5} & 370.7 & 404.6 & 96.4 & 285.0 & 353.8 & \textbf{372.8} & 351.3 & 368.8 & 70.4 \\
\midrule
\multirow{3}{*}{\syn} & 100 & \textbf{97.2} & 95.6 & 96.3 & 97.0 & 97.1 & 56.3 & 89.4 & 86.8 & 96.5 & 95.0 & \textbf{98.6} & 64.8 & 90.8 & 88.7 & 95.5 & \textbf{96.0} & 87.6 & 53.7 \\
 & 300 & 292.1 & 287.2 & \textbf{292.9} & 290.0 & 292.2 & 174.3 & 266.4 & 267.5 & 287.8 & 290.5 & \textbf{294.4} & 196.9 & 264.1 & 265.4 & 286.2 & \textbf{290.4} & 273.9 & 159.4 \\
 & 500 & 487.6 & 483.3 & \textbf{489.0} & 482.2 & 488.6 & 293.5 & 444.5 & 451.9 & 481.3 & 484.4 & \textbf{486.0} & 328.6 & 438.3 & 446.3 & 477.0 & \textbf{486.6} & 462.9 & 269.6 \\
\midrule
\multirow{3}{*}{\topic} & 100 & 51.7 & 83.4 & 89.8 & 93.7 & \textbf{96.1} & 15.0 & 56.4 & 69.9 & 76.4 & \textbf{86.9} & 84.4 & 20.4 & 52.0 & 66.2 & 79.4 & \textbf{91.1} & 89.8 & 22.5 \\
 & 300 & 124.0 & 214.5 & 242.4 & \textbf{266.1} & 262.7 & 49.0 & 133.5 & 185.2 & 208.4 & \textbf{246.4} & 233.9 & 62.8 & 125.1 & 172.4 & 211.6 & \textbf{253.5} & 251.0 & 69.2 \\
 & 500 & 182.6 & 324.9 & 360.9 & \textbf{419.8} & 403.9 & 83.1 & 199.7 & 293.2 & 334.6 & \textbf{393.7} & 368.5 & 105.7 & 190.5 & 266.2 & 328.4 & \textbf{397.4} & 391.0 & 115.1 \\
\midrule
\multirow{3}{*}{\mbpp} & 100 & 51.3 & 56.8 & \textbf{70.9} & 56.8 & 64.4 & 34.0 & 57.7 & 68.1 & \textbf{83.6} & 68.6 & 81.4 & 36.5 & 39.4 & 50.2 & \textbf{62.4} & 47.1 & 58.5 & 25.0 \\
 & 300 & 120.2 & 122.7 & \textbf{138.7} & 117.5 & 133.2 & 104.9 & 129.2 & 141.3 & \textbf{164.0} & 134.3 & 146.8 & 108.9 & 92.4 & 93.3 & \textbf{113.5} & 78.6 & 98.5 & 83.0 \\
 & 500 & -- & -- & -- & -- & -- & -- & -- & -- & -- & -- & -- & -- & -- & -- & -- & -- & -- & -- \\
\midrule
\multirow{3}{*}{\live} & 100 & 86.6 & 91.2 & \textbf{96.6} & 89.2 & 84.6 & 74.2 & 99.4 & 97.3 & \textbf{99.5} & 92.2 & 95.1 & 81.0 & 86.3 & 91.3 & \textbf{94.9} & 71.2 & 79.4 & 54.7 \\
 & 300 & 240.2 & 268.6 & \textbf{276.7} & 251.8 & 244.1 & 221.6 & 279.9 & 280.4 & \textbf{291.0} & 273.0 & 282.3 & 246.1 & 208.1 & 241.3 & \textbf{255.1} & 184.5 & 198.5 & 168.4 \\
 & 500 & -- & -- & -- & -- & -- & -- & -- & -- & -- & -- & -- & -- & -- & -- & -- & -- & -- & -- \\
\bottomrule
\end{tabular}%
}
\end{table*}

\subsection{RQ2: Out-of-distribution Input Detection}
\label{sec:rq2}

Before using a predicted failure risk to prioritise an input, a tester may ask whether the labelled references contain comparable inputs. RQ2 tests whether \name's dual GMM marginal reference density identifies inputs drawn from other tasks, and whether the LR can provide the same signal.

Across every target--source pair, reference density achieves lower mean FPR95 and higher mean ROC-AUC than either LR-derived score (Table~\ref{tab:rq2-ood}). Averaged over the four OOD sources, its FPR95 is $0.185$ for \json and $0.164$ for \topic, compared with $0.879$ and $0.639$ for pass probability. At these thresholds, reference density retains $94.7\%$ and $95.2\%$ of held-out in-distribution inputs, close to the $95\%$ calibration target. The practical implication is clear. At thresholds that retain $95\%$ of ID inputs, LR pass probability accept most OOD inputs as in-distribution; maximum-softmax confidence accepts over $95\%$ of them for both targets. A high pass probability, or softmax confidence, therefore does not indicate that an input belongs to the reference task, and neither score can serve as a reliable OOD detection filter under this protocol.

Detection difficulty nonetheless varies across sources, which may partly reflect characteristics shared with the target task. For \json, FPR95 is relatively high on \syn inputs, possibly because JSON and source code share syntactic structure. For \topic, the higher FPR95 on \model inputs may reflect the expository prose common to abstracts and news text, which may partially explain the difficulty to separate them.

\begin{rqbox}[Answer to RQ2]
\name's dual GMM reference-density score detects the evaluated cross-task inputs substantially more effectively than LR-derived pass probability or confidence, outperforming both on every target--source pair while retaining approximately $95\%$ of held-out in-distribution inputs. Detection effectiveness nonetheless remains source-dependent.
\end{rqbox}

\begin{table}[t]
\caption{Per-source and average out-of-distribution detection results.}
\label{tab:rq2-ood}
\footnotesize
\resizebox{\columnwidth}{!}{%
  \begin{tabular}{lcc|lcc|lcc|lcc}
  \toprule
  \multicolumn{6}{c}{Dual GMM} & \multicolumn{6}{c}{LR} \\ \midrule
  Target & \multicolumn{2}{c|}{\json} & Target & \multicolumn{2}{c|}{\topic} & Target & \multicolumn{2}{c|}{\json} & Target & \multicolumn{2}{c}{\topic} \\
  \midrule
  OOD Source & FPR95 & ROC-AUC & OOD Source & FPR95 & ROC-AUC & OOD Source & FPR95 & ROC-AUC & OOD Source & FPR95 & ROC-AUC \\
  \midrule
  \topic  & 0.342 $\pm$ 0.28 & 0.940 $\pm$ 0.05 & \json   & 0.004 $\pm$ 0.00 & 0.999 $\pm$ 0.00 & \topic  & 0.812 $\pm$ 0.25 & 0.629 $\pm$ 0.23 & \json   & 0.372 $\pm$ 0.26 & 0.861 $\pm$ 0.13 \\
  \spell  & 0.172 $\pm$ 0.11 & 0.973 $\pm$ 0.02 & \spell  & 0.196 $\pm$ 0.07 & 0.965 $\pm$ 0.01 & \spell  & 0.961 $\pm$ 0.05 & 0.441 $\pm$ 0.20 & \spell  & 0.353 $\pm$ 0.13 & 0.871 $\pm$ 0.06 \\
  \syn    & 0.222 $\pm$ 0.15 & 0.952 $\pm$ 0.03 & \syn    & 0.001 $\pm$ 0.00 & 0.999 $\pm$ 0.00 & \syn    & 0.782 $\pm$ 0.30 & 0.581 $\pm$ 0.26 & \syn    & 0.874 $\pm$ 0.11 & 0.411 $\pm$ 0.23 \\
  \model  & 0.005 $\pm$ 0.00 & 0.999 $\pm$ 0.00 & \model  & 0.455 $\pm$ 0.11 & 0.892 $\pm$ 0.03 & \model  & 0.959 $\pm$ 0.07 & 0.227 $\pm$ 0.24 & \model  & 0.958 $\pm$ 0.04 & 0.551 $\pm$ 0.23 \\
  Average & 0.185 $\pm$ 0.13 & 0.966 $\pm$ 0.02 & Average & 0.164 $\pm$ 0.03 & 0.964 $\pm$ 0.00 & Average & 0.879 $\pm$ 0.10 & 0.469 $\pm$ 0.18 & Average & 0.639 $\pm$ 0.11 & 0.674 $\pm$ 0.10 \\
  \bottomrule
  \end{tabular}%
}
\end{table}

\subsection{RQ3: Characterisation of Failure Patterns}
\label{sec:rq3}

For failing-class component $j$, we select inputs whose posterior failure probability and responsibility within the failing-class GMM for $j$ are both at least 0.95, then retain those observed to fail under the evaluation's pass/fail criterion. We denote the resulting set by $F_j$. 

\subsubsection{RQ3-1: Failure Patterns Across Components}
\label{sec:rq3-1} 

\begin{figure}[t]
\centering
\includegraphics[width=\linewidth]{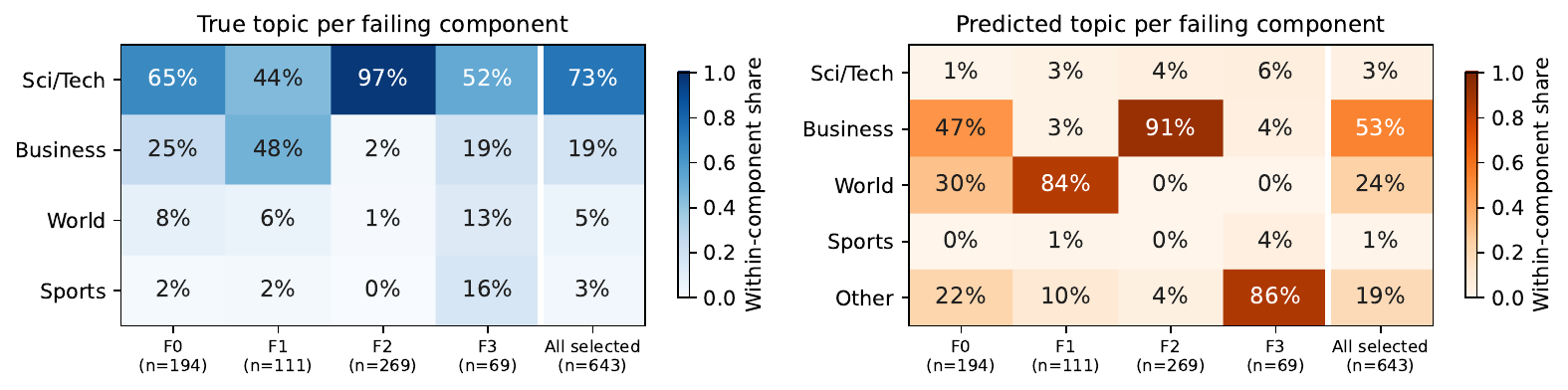}
\caption{True-topic (left) and predicted-topic (right) distributions among selected failures in each failing-class component of \name on \topic with Llama3-8B (seed~0).}
\label{fig:rq3-1_heatmap}
\end{figure}

Fig.~\ref{fig:rq3-1_heatmap} shows the true-topic (left) and predicted-topic (right) distributions of components $F_0$--$F_3$ for seed~0, with the final column of each panel giving the pooled distribution across all four components. Read together, the panels show different failure patterns. Sci/Tech accounts for 73\% of the failures in these sets and appears in several components, but the components' true--predicted topic differ. $F_2$ isolates a single Sci/Tech$\rightarrow$Business confusion, whereas $F_1$ gathers inputs from several true topics that the model mispredicts as World, and $F_3$ concentrates on inputs mispredicted as Other. $F_0$, by contrast, remains mixed and shows no dominant pattern. Combined, these patterns suggest that the components capture variations in how inputs are misclassified, although distinct behaviour may not be evident in every component.

Component $F_2$ concentrates on Sci/Tech$\rightarrow$Business errors: 97\% of its inputs have ground-truth topic Sci/Tech but 91\% are predicted as Business. $F_1$ mixes Business (48\%) and Sci/Tech (44\%) inputs, but 84\% of its predictions are World. In $F_3$, 86\% of predictions fall outside the four topic labels (``Other''). $F_0$ has more varied predictions: Business (47\%), World (30\%), and Other (22\%) but 65\% are actually labelled Sci/Tech.

Across ten seeds, the dual GMM identifies $3.3 \pm 1.1$ failing-class components on average. We measure association using Cram\'er's $V$, where 0 means no association and 1 means maximal association. Component membership is associated with true--predicted topic pairs ($V=0.71 \pm 0.07$), more strongly with predicted topics ($0.67 \pm 0.08$) than with true topics ($0.40 \pm 0.07$). Thus, knowing an input's component gives information about its observed error, particularly which label the LLM predicts. A failure score can only rank inputs by predicted risk; the failing-class components additionally group failures by recurring error patterns, although some remain mixed.

\begin{table}[t]
\centering
\caption{Failing-class component tokens on \topic and the top 20 tokens with positive (LR$^{+}$, failure) or negative (LR$^{-}$, passing) LR probe weights.}
\label{tab:rq3-topic}
\footnotesize
\resizebox{0.9\columnwidth}{!}{%
  \begin{tabular}{lr>{\raggedright\arraybackslash}p{0.85\columnwidth}}
  \toprule
  & $|F_j|$ & Top-20 tokens \\
  \midrule
  $F_0$ & 194 & \textit{law, legal, legality, laws, legally, legislation, legislative, lawyers, juris, lawyer, lawsuits, legis, jur, judicial, legislature, government, courtroom, justice, legisl, regulatory} \\
  \addlinespace
  $F_1$ & 111 & \textit{global, international, globally, worldwide, geopol, globalization, geopolitical, internationally, internacional, globe, diplomatic, world, geography, glob, ernational, diplomacy, diplom, unesco, globals, europe} \\
  \addlinespace
  $F_2$ & 269 & \textit{companies, company, tech, business, businesses, technology, corporations, corporate, software, microsoft, industry, computer, industries, empresas, corporation, empresa, ibm, marketing, computers, manufacturers} \\
  \addlinespace
  $F_3$ & 69 & \textit{entertainment, celebrity, celebrities, personality, cultural, persona, hobby, fandom, person, personalities, recreational, personas, personen, musician, individuals, science, history, human, humanity, americans} \\
  \midrule
  LR$^{+}$ & -- & \textit{mvp, wereld, cultural, etsy, climate, locals, entertainment, administration, countryside, vendor, lawsuits, globals, china, lobbyist, healthcare, venture, travel, incidents, salesforce, nokia} \\
  \addlinespace
  LR$^{-}$ & -- & \textit{nationally, wrestling, juventus, uruguay, ski, asia, yemen, \begin{CJK*}{UTF8}{gbsn}科学\end{CJK*}[science], surname, african, terms, ge, \begin{CJK*}{UTF8}{mj}산업\end{CJK*}[industry], products, automotive, otechnology, retail, monarch, brokerage, corp} \\
  \bottomrule
  \end{tabular}%
}
\end{table}
\subsubsection{RQ3-2: Semantic Characterisation}
\label{sec:rq3-2}

True and predicted topics describe the observed failures but J-lens profiles can add descriptions in vocabulary terms. Table~\ref{tab:rq3-topic} lists the tokens whose average J-lens scores are most increased among each component's failures, compared with their averages across all inputs: legal terms in $F_0$, international terms in $F_1$, corporate and technology terms in $F_2$, and celebrity and entertainment terms in $F_3$. As we saw above, both $F_0$ and $F_2$ contain Sci/Tech articles predicted as Business, but their profiles distinguish legal ($F_0$) from corporate and technology concepts ($F_2$). The readout tokens can thus distinguish groups that share a failure pattern but not the same emphasised terms.

\begin{figure}[t]
    \centering
    \begin{subfigure}[t]{0.49\linewidth}
        \centering
        \includegraphics[width=\linewidth]{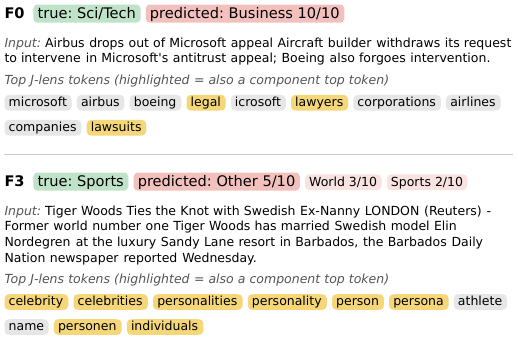}
        \caption{\topic}
        \label{fig:rq3/rq3_topic}
    \end{subfigure}\hfill
    \begin{subfigure}[t]{0.49\linewidth}
        \centering
        \includegraphics[width=\linewidth]{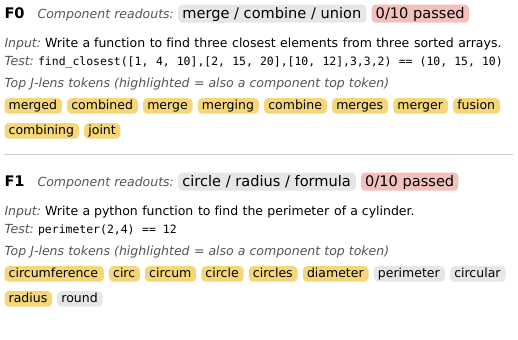}
        \caption{\mbpp}
        \label{fig:rq3/rq3_mbpp}
    \end{subfigure}
    \caption{Representative failures of the failing-class components}
    \label{fig:rq3-cases}
\end{figure}

JRS can also make LR weights readable as tokens. LR learns a single scoring direction to separate passing from failing inputs, so its global weight ranking combines token contributions across different failure groups. The inspected positive and negative weight rankings (last rows of Table~\ref{tab:rq3-topic}) consequently mix several concepts. The GMM component profiles give more focused descriptions, including differences between groups that share a label confusion.

At the input level, the readouts give finer details. Fig.~\ref{fig:rq3/rq3_topic} shows one failing input from each of two components of \topic, $F_0$ and $F_3$, with its leading tokens. In the $F_0$ article about Airbus withdrawing from Microsoft's antitrust appeal, the readout includes company names alongside legal (\textit{legal, lawyers, lawsuits}) and corporate (\textit{corporations, companies}) terms. All ten responses predict Business, although the article ground truth label is Sci/Tech. In the Sports-labelled article about Tiger Woods' wedding, the $F_3$ readout includes \textit{athlete} but is led by terms like \textit{celebrity} and \textit{personalities}; five of ten responses predict Other.

Reading Microsoft's antitrust appeal as Business, or a golfer's wedding as celebrity news, is defensible. These cases suggest possible ambiguity in the category boundaries and ground-truth labels, rather than necessarily a misunderstanding of the articles by the LLM. Together with the inputs and generated responses, the readouts can help an analyst identify plausible alternative interpretations that a failure score alone does not describe.

The value of J-lens readouts also extends to program synthesis. In Fig.~\ref{fig:rq3/rq3_mbpp}, on the \mbpp task, $F_0$ emphasises merging (\textit{merge, combine}), while $F_1$ emphasises circle geometry (\textit{circumference, radius}). In the $F_0$ case, the task asks for the three closest elements, one from each of three sorted arrays. All ten leading J-lens tokens concern merging or combining, but a majority of the responses merge the arrays without enforcing the one-per-array constraint; none passes. In the $F_1$ case, the task asks for the perimeter of a cylinder. The benchmark expects the perimeter of its side-view rectangle, $2(d+h)$, but the task does not say so. The leading J-lens tokens are circle terms (\textit{circumference, circle, radius}), with no rectangle term among them, and all ten responses build circle-based formulas with $\pi$; none passes. The readout and generated programs together suggest an alternative interpretation of an underspecified task.

These cases show how pre-generation readouts can help characterise failures and identify possible ambiguity in task specifications. The cylinder's perimeter has no standard definition, making an alternative interpretation defensible. As with the golfer's wedding in \topic, the readout and generated responses together help an analyst assess whether the failure reflects an LLM error or ambiguity in the task specification or ground-truth label. However, the readouts do not establish that the highlighted concepts caused the failures. A readout may also suggest an appropriate approach without revealing whether the generated program will implement it correctly.

\begin{rqbox}[Answer to RQ3]
\name helps characterise failure groups and individual cases through J-lens token profiles. On \topic, GMM component profiles distinguish concepts even between groups sharing a label confusion. Individual readouts on \topic and \mbpp can suggest ambiguity in labels or task specifications. LR weights are also token-readable, but their global rankings mix concepts rather than providing group-specific profiles.
\end{rqbox}

\begin{table}[ht]
\centering
\caption{Normalised \fan on each target model, grouped by the open-weight source model used for scoring, averaged over ten tasks and ten seeds.}
\label{tab:transfer-failure25}
\resizebox{\columnwidth}{!}{%
\begin{tabular}{llrrrrr|llrrrrr}
\toprule
Target & Source & Clotho & LIHS$_{\mathrm{DG}}$ & LIHS$_{\mathrm{LR}}$ & JRS$_{\mathrm{DG}}$ & JRS$_{\mathrm{LR}}$ & Target & Source & Clotho & LIHS$_{\mathrm{DG}}$ & LIHS$_{\mathrm{LR}}$ & JRS$_{\mathrm{DG}}$ & JRS$_{\mathrm{LR}}$ \\
\midrule
\multirow{3}{*}{Llama-3.3-70B} & Llama-3.1-8B & 0.546 & 0.562 & \textbf{0.614} & 0.545 & \underline{0.610} & \multirow{3}{*}{GPT-4o-mini} & Llama-3.1-8B & 0.515 & 0.547 & \underline{0.597} & 0.564 & \textbf{0.607} \\
 & Gemma-2-9B & 0.483 & 0.574 & \textbf{0.634} & 0.586 & \underline{0.626} &  & Gemma-2-9B & 0.477 & 0.590 & \underline{0.644} & 0.591 & \textbf{0.648} \\
 & Qwen3-8B & 0.488 & 0.574 & \underline{0.621} & 0.577 & \textbf{0.633} &  & Qwen3-8B & 0.481 & 0.590 & \underline{0.662} & 0.622 & \textbf{0.676} \\\midrule
\multirow{3}{*}{Gemma-3-27B} & Llama-3.1-8B & 0.522 & 0.573 & \textbf{0.617} & 0.581 & \underline{0.611} & \multirow{3}{*}{Claude Haiku 4.5} & Llama-3.1-8B & 0.479 & 0.498 & \textbf{0.537} & 0.475 & \underline{0.528} \\
 & Gemma-2-9B & 0.520 & 0.606 & \underline{0.664} & 0.619 & \textbf{0.673} &  & Gemma-2-9B & 0.415 & 0.492 & \underline{0.553} & 0.502 & \textbf{0.556} \\
 & Qwen3-8B & 0.482 & 0.580 & \underline{0.652} & 0.613 & \textbf{0.656} &  & Qwen3-8B & 0.435 & 0.513 & \underline{0.582} & 0.533 & \textbf{0.587} \\\midrule
\multirow{3}{*}{Qwen3-32B} & Llama-3.1-8B & 0.494 & 0.516 & \textbf{0.549} & 0.493 & \underline{0.549} & \multirow{3}{*}{Gemini 3.5 Flash Lite} & Llama-3.1-8B & 0.453 & 0.481 & \textbf{0.517} & 0.474 & \underline{0.508} \\
 & Gemma-2-9B & 0.432 & 0.515 & \textbf{0.579} & 0.531 & \underline{0.572} &  & Gemma-2-9B & 0.447 & 0.519 & \textbf{0.587} & 0.535 & \underline{0.586} \\
 & Qwen3-8B & 0.455 & 0.533 & \underline{0.578} & 0.541 & \textbf{0.604} &  & Qwen3-8B & 0.430 & 0.512 & \textbf{0.588} & 0.548 & \underline{0.585} \\
\bottomrule
\end{tabular}}
\end{table}

\subsection{RQ4: Transferability of Failure Predictions}
\label{sec:rq4}

Table~\ref{tab:transfer-failure25} reports normalised \fan. A value of 0.6 means that the first quarter of an ordering contains 60\% as many failures as a perfect ordering would place there. Across the six targets, Clotho's pass-only score~\cite{yoon2026clotho} averages 0.48. Both LR probes exceed Clotho on every source--target pair, averaging 0.60, a gain of 0.12. The dual GMM scorers average 0.55, a gain of 0.07 that depends more on the source: 0.09 for Gemma-2-9B and 0.10 for Qwen3-8B, but 0.02 for Llama-3.1-8B. Scorer choice has the larger effect. With the representation fixed, LR exceeds dual GMM in all 36 comparisons, by 0.05 on average. With the scorer fixed, LIHS and J-lens differ by only 0.01 on average, with J-lens ahead in 23 of the 36.

The gains extend to all three proprietary targets. Claude Haiku 4.5 and Gemini 3.5 Flash Lite pass most inputs, with mean pass rates of 0.857 and 0.860 across tasks. Even on these two targets, all twelve LR scores exceed their source-matched Clotho scores; their mean normalised \fan is 0.559 versus 0.443. The larger open-weight targets also let us assess the role of model family. We compare each source's lead over the other two sources on its own-family target with its lead on the other open-weight targets. Sharing a family gives an average advantage of about 0.02, much smaller than the LR probes' gain over Clotho. It does not overcome source quality: Llama-3.1-8B is the weakest source in 22 of the 24 target--scorer combinations, including all four combinations in Llama-3.3-70B.

\begin{rqbox}[Answer to RQ4]
Scores from smaller open-weight models can prioritise failures in larger and proprietary models better than Clotho. The LR scores transfer the best, improving normalised \fan over Clotho on every source--target pair by 0.12 on average; on GPT-4o-mini, J-lens LR from Qwen3-8B reaches 0.676, against 0.481 for Clotho. The effectiveness of transfer is slightly better within the same model family, but specific source and target pairs matter more.
\end{rqbox}

\subsection{RQ5: Computational Cost}
\label{sec:rq5} 

\begin{figure}[ht]
    \centering
    \begin{subfigure}[t]{0.51\linewidth}
        \vspace{0pt}
        \centering
        \parbox[c][2.6cm][t]{\linewidth}{%
            \centering
            \includegraphics[width=\linewidth,height=3cm,keepaspectratio]
                {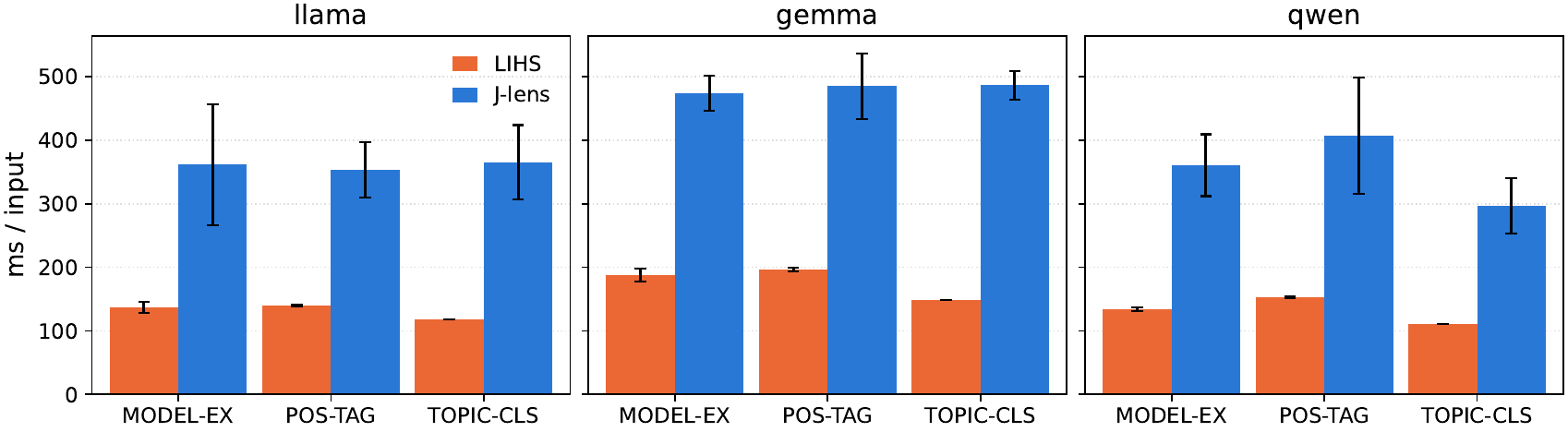}%
        }
        \caption{Average extraction time per input}
        \label{fig:rq5/rq5_extraction}
    \end{subfigure}\hfill
    \begin{subfigure}[t]{0.49\linewidth}
        \vspace{0pt}
        \centering
        \parbox[c][2.6cm][t]{\linewidth}{%
            \centering
            \includegraphics[width=\linewidth,height=3cm,keepaspectratio]
                {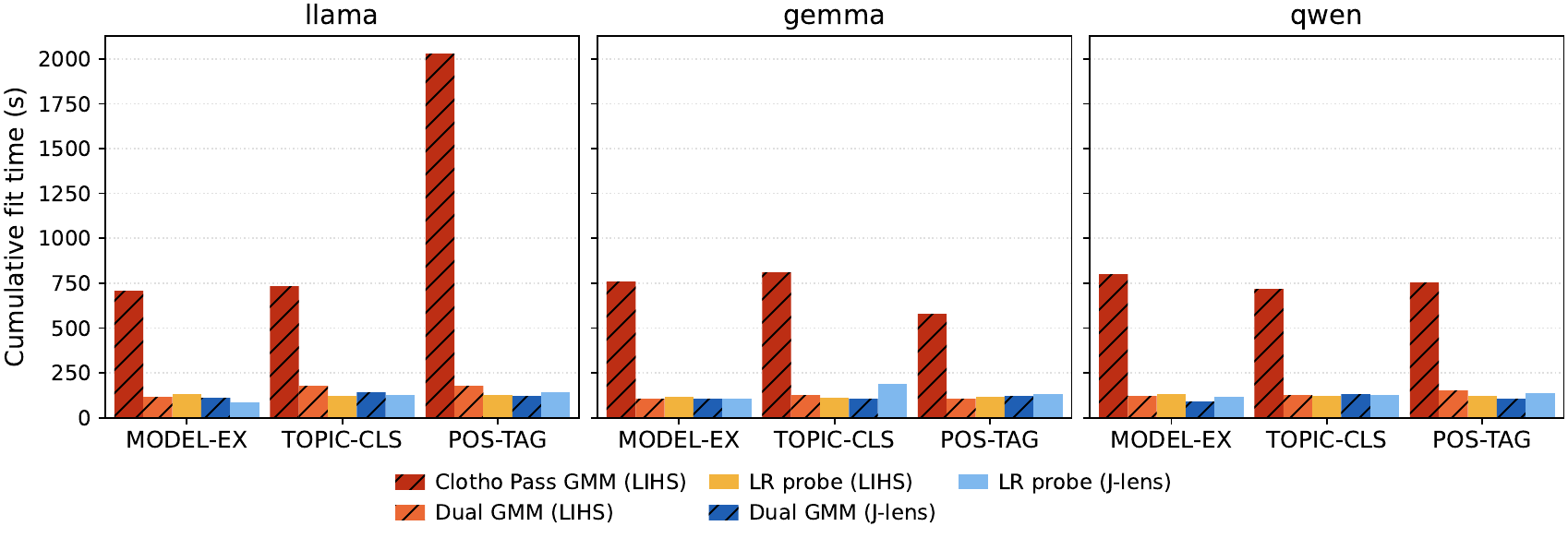}%
        }
        \caption{Cumulative fitting time}
        \label{fig:rq5/rq5_fitting}
    \end{subfigure}
    \caption{Computational cost of representation extraction and scorer fitting for each task--model combination.}
    \label{fig:rq5-cost}
\end{figure}

Figure~\ref{fig:rq5/rq5_extraction} shows extraction time per input. J-lens takes 0.40\,s per input on average, compared with 0.15\,s for LIHS, or 2.7 times as long. This is expected, as LIHS is the hidden state read directly from the forward pass, whereas J-lens must further map that state into vocabulary space through the Jacobian lens~\cite{gurnee2026verbalizable}. Gemma, the largest of the three studied LLMs, is the slowest against both representations, yet J-lens extraction stays below half a second per input in every setting on a single consumer GPU.

Figure~\ref{fig:rq5/rq5_fitting} shows the cumulative fitting time over the 500-label active-learning loop. Clotho is by far the slowest: fitting its passing-class GMM takes 874\,s per run on average, about seven times the roughly two-minute average of the other four pairs. We attribute the difference mainly to how the covariance is modelled. Both GMM-based methods fit PCA-reduced features, but Clotho estimates a full covariance matrix for each mixture component, whereas the dual GMM usually falls back to diagonal covariance because each class has too few references to estimate a full one.

Among these four pairs, the representation has a small, scorer-dependent effect: J-lens lowers the dual GMM's average fitting time from 133\,s to 113\,s (15\%) but raises the LR probe's from 122\,s to 131\,s (7\%). The scorer likewise has a small, representation-dependent effect: replacing the dual GMM with the LR probe lowers the average from 133\,s to 122\,s on LIHS (8\%) but raises it from 113\,s to 131\,s on J-lens (16\%). As a result, the dual GMM on J-lens is the cheapest of the five pairs on average. The two scorers have comparable costs even though a GMM is the more complex model, because both spend most of their time on hyperparameter search: the dual GMM runs about 500 small fits per iteration on PCA-reduced features, whereas the LR probe runs about 50 L1-regularised fits on all original feature space. This parity is specific to RQ1's search configuration rather than a general property of GMMs versus logistic regression.

\begin{rqbox}[Answer to RQ5]
\name's additional cost lies in extraction: J-lens adds about 0.25\,s per input over LIHS, yet stays below 0.5\,s per input on a consumer GPU. On J-lens, the dual GMM and the LR probe each take about two minutes to fit over the 500-label loop, whereas Clotho's passing-class GMM takes 7.7$\times$ as long as the dual GMM and 6.7$\times$ as long as the LR probe.
\end{rqbox}

\section{Discussion}
\label{sec:discussion}
\noindent\textbf{What J-lens contributes.}
Derived from the same forward pass as LIHS, JRS adds a readable basis for characterising failures rather than new information. Its comparable average discrimination suggests little predictive cost for this readability. Task differences remain: JRS$_\mathrm{LR}$ leads on \topic and \pos, while LIHS$_\mathrm{LR}$ leads on \mbpp and \live. These differences may reflect how the corpus used to estimate the Jacobians and token selection shape the task-relevant information JRS retains. Varying this corpus, vocabulary size, and layer range could assess sensitivity to these choices.

J-lens readouts are essentially decoded internal layer activations. Consequently, while some may support failure characterisation, fundamentally they lack any self-reflective diagnosis capabilities. For example, we observe that J-lens readouts correctly point to dynamic programming as the approach towards the solution for \mbpp, only to fail to generate a correct implementation. Further studies are required to explore whether self-diagnostic prompting can elicit deeper characterisation.

Component profiles characterise failure groups through prominent tokens, while individual readouts can suggest alternative interpretations of inputs or task specifications. These descriptions come at the extra extraction cost reported in Section~\ref{sec:rq5}. Readouts could also characterise LIHS-fitted groups; whether fitting in JRS yields more coherent groups remains untested. Our descriptive interpretation is limited to a few cases. Future studies, potentially involving human users, should assess whether readouts help analysts characterise failures and revise task prompts beyond what inputs and generated responses alone support.

\noindent\textbf{Concept-level test coverage.} \name characterises groups of failing inputs through their prominent readout tokens. Such group-level summaries may obscure the multiple concepts within an individual input, as the Airbus article illustrates (Section~\ref{sec:rq3-2}). Future work could jointly identify shared concepts and overlapping groups of inputs, allowing an input to belong to several groups. Concept combinations could then serve as discrete, interpretable units of test coverage. Testers could prioritise combinations associated with frequent failures and generate inputs for combinations poorly represented in the test set. Defining the relevant concept space and establishing criteria for adequate coverage remain open questions.

\section{Threats to Validity}
\label{sec:threats}

Threats to internal validity concern anything that could interfere with the observed effect of \name, namely its predictive and characterising power. JRS depends on subsets of tokens the LLM is ready to verbalise, hence we cannot rule out possibilities that certain tokens are correlated with failures without actually causing them. We try to minimise such threats by presenting the actual readout tokens in RQ3 for a qualitative analysis. However, only a full understanding of LLM internals can fundamentally rid us of such threats. External validity lies on the degree of generalisation of the reported results. Our study primarily reports on J-lenses for three 8-9B open weight LLMs, against ten tasks; similarly, the transfer study only concerns three proprietary models. Other architectures, such as reasoning models, may produce different Jacobian workspaces, resulting in different results. RQ2 also operationalises distribution shift using other tasks' inputs placed into the target prompt; such inputs may be easier to detect than drift encountered in deployment. Only further experimentation can reduce threats to external validity. Finally, threats to construct validity concerns whether we are measuring what we intend to observe. We use established measures (ROC-AUC, Failure@$N$, FPR95), as well as \texttt{scikit-learn} implementations that have withstood public scrutiny, to mitigate threats. However, these measures fundamentally remain proxies. 

\section{Conclusion}
\label{sec:conclusion}
We propose \name, a pre-generation test adequacy framework for LLMs that learns from both passing and failing reference inputs. \name represents each input either as hidden states or as vocabulary-aligned J-lens readouts. Across ten tasks and three open-weight LLMs, \name outperforms Clotho, the existing state-of-the-art pre-generation adequacy metric. The LR probe performs the best for failure prediction, with the average ROC-AUC 0.84 over 0.68 of Clotho; the dual GMM provides a reference density, which can flag out-of-distribution inputs with much lower false positive rates than LR. Most importantly, dual GMMs can group failures that share the same confusion; further, J-lens readouts can characterise those failures semantically. Finally, scores computed on small models transfer to larger and proprietary targets, raising normalised \fan from 0.48 to 0.60, allowing testers to prioritise inputs for expensive models before querying them. In future work, we plan to build on these concept-level groupings to measure concept-level test coverage and to generate inputs for poorly covered concept combinations.



\bibliographystyle{ACM-Reference-Format}
\bibliography{references}

@misc{replication,
	author = {Anonymous},
	howpublished = {\url{https://anonymous.4open.science/r/foreword-3}},
	note = {Accessed: 2026-10-03},
	title = {Replication Package for Foreword},
	year = {2026}}

@book{cramer1946mathematical,
 author    = {Cram{\'e}r, Harald},
 title     = {Mathematical Methods of Statistics},
 year      = {1946},
 publisher = {Princeton University Press},
 address   = {Princeton, NJ}
}

@inproceedings{Zhou2025aa,
	address = {Los Alamitos, CA, USA},
	author = {Zhou, Shide and Li, Tianlin and Wang, Kailong and Huang, Yihao and Shi, Ling and Liu, Yang and Wang, Haoyu},
	booktitle = {2025 IEEE/ACM 47th International Conference on Software Engineering (ICSE)},
	doi = {10.1109/ICSE55347.2025.00209},
	month = May,
	pages = {730-742},
	publisher = {IEEE Computer Society},
	title = {{ Understanding the Effectiveness of Coverage Criteria for Large Language Models: A Special Angle from Jailbreak Attacks }},
	url = {https://doi.ieeecomputersociety.org/10.1109/ICSE55347.2025.00209},
	year = {2025}}

@inproceedings{Azaria2023aa,
	address = {Singapore},
	author = {Azaria, Amos and Mitchell, Tom},
	booktitle = {Findings of the Association for Computational Linguistics: EMNLP 2023},
	doi = {10.18653/v1/2023.findings-emnlp.68},
	editor = {Bouamor, Houda and Pino, Juan and Bali, Kalika},
	month = dec,
	pages = {967--976},
	publisher = {Association for Computational Linguistics},
	title = {The Internal State of an {LLM} Knows When It{'s} Lying},
	url = {https://aclanthology.org/2023.findings-emnlp.68/},
	year = {2023}}

@article{Zhu1997aa,
	address = {New York, NY, USA},
	author = {Zhu, Hong and Hall, Patrick A. V. and May, John H. R.},
	doi = {10.1145/267580.267590},
	issn = {0360-0300},
	issue_date = {Dec. 1997},
	journal = {ACM Comput. Surv.},
	month = dec,
	number = {4},
	numpages = {62},
	pages = {366--427},
	publisher = {Association for Computing Machinery},
	title = {Software unit test coverage and adequacy},
	url = {https://doi.org/10.1145/267580.267590},
	volume = {29},
	year = {1997}}

@inproceedings{Elbaum:2001lr,
	author = {Sebastian G. Elbaum and Alexey G. Malishevsky and Gregg Rothermel},
	booktitle = {Proceedings of the International Conference on Software Engineering (ICSE 2001)},
	month = {May},
	pages = {329-338},
	publisher = {{ACM} Press},
	title = {Incorporating Varying Test Costs and Fault Severities into Test Case Prioritization},
	year = {2001}}

@article{Farquhar2024aa,
	author = {Farquhar, Sebastian and Kossen, Jannik and Kuhn, Lorenz and Gal, Yarin},
	date = {2024/06/01},
	doi = {https://doi.org/10.1038/s41586-024-07421-0},
	id = {Farquhar2024},
	isbn = {1476-4687},
	journal = {Nature},
	number = {8017},
	pages = {625--630},
	title = {Detecting hallucinations in large language models using semantic entropy},
	volume = {630},
	year = {2024}}

@inproceedings{Manakul2023aa,
	address = {Singapore},
	author = {Manakul, Potsawee and Liusie, Adian and Gales, Mark},
	booktitle = {Proceedings of the 2023 Conference on Empirical Methods in Natural Language Processing},
	doi = {10.18653/v1/2023.emnlp-main.557},
	editor = {Bouamor, Houda and Pino, Juan and Bali, Kalika},
	month = {December},
	pages = {9004--9017},
	publisher = {Association for Computational Linguistics},
	title = {{S}elf{C}heck{GPT}: Zero-Resource Black-Box Hallucination Detection for Generative Large Language Models},
	year = {2023}}

@inproceedings{Tan2025aa,
	author = {Tan, Sijun and Zhuang, Siyuan and Montgomery, Kyle and Tang, William and Cuadron, Alejandro and Wang, Chenguang and Popa, Raluca and Stoica, Ion},
	booktitle = {International Conference on Learning Representations},
	pages = {63277--63303},
	title = {Judgebench: A benchmark for evaluating llm-based judges},
	volume = {2025},
	year = {2025}}

@inproceedings{Agarwal2026aa,
	author = {Agarwal, Shyam and He, Hao and Vasilescu, Bogdan},
	booktitle = {Proceedings of the 23rd International Conference on Mining Software Repositories},
	isbn = {9798400724749},
	numpages = {6},
	pages = {857--862},
	title = {{AI IDEs} or Autonomous Agents? Measuring the Impact of Coding Agents on Software Development},
	doi = {https://doi.org/10.1145/3793302.3793589},
	year = {2026}}

@article{Qiu2024aa,
	author = {Qiu, Jianing and Lam, Kyle and Li, Guohao and Acharya, Amish and Wong, Tien Yin and Darzi, Ara and Yuan, Wu and Topol, Eric J},
	journal = {Nature Machine Intelligence},
	number = {12},
	pages = {1418--1420},
	publisher = {Nature Publishing Group UK London},
	title = {LLM-based agentic systems in medicine and healthcare},
	volume = {6},
	doi = {https://doi.org/10.1038/s42256-024-00944-1},
	year = {2024}}

@article{Chu2025aa,
	author = {Chu, Zhendong and Wang, Shen and Xie, Jian and Zhu, Tinghui and Yan, Yibo and Ye, Jingheng and Zhong, Aoxiao and Hu, Xuming and Liang, Jing and Philip, S Yu and others},
	journal = {EMNLP (Findings)},
	pages = {13782--13810},
	title = {LLM Agents for Education: Advances and Applications.},
	doi = {https://doi.org/10.18653/v1/2025.findings-emnlp.743},
	year = {2025}}

@article{yoon2026clotho,
	author = {Yoon, Juyeon and Kim, Somin and Feldt, Robert and Yoo, Shin},
	journal = {Proceedings of the ACM on Software Engineering},
	number = {FSE},
	pages = {1910--1932},
	publisher = {ACM New York, NY, USA},
	title = {Clotho: Measuring task-specific pre-generation test adequacy for {LLM} inputs},
	volume = {3},
	doi = {https://doi.org/10.1145/3797114},
	year = {2026}}

@article{dobslaw2026taxonomy,
	author = {Dobslaw, Felix and Feldt, Robert and Yoon, Juyeon and Yoo, Shin},
	doi = {10.1145/3806396},
	journal = {ACM Transactions on Software Engineering and Methodology},
	number = {9},
	pages = {261:1--261:24},
	title = {Challenges in Testing Large Language Model-Based Software: A Faceted Taxonomy},
	volume = {35},
	year = {2026}}

@article{gurnee2026verbalizable,
	author = {Gurnee, Wes and Sofroniew, Nicholas and Pearce, Adam and Piotrowski, Mateusz and Kauvar, Isaac and Chen, Runjin and Soligo, Anna and Bogdan, Paul and Ong, Euan and Wang, Rowan and others},
	journal = {arXiv preprint arXiv:2607.15495},
	title = {Verbalizable representations form a global workspace in language models},
	year = {2026}}

@article{hagiwara2019github,
	author = {Hagiwara, Masato and Mita, Masato},
	journal = {arXiv preprint arXiv:1911.12893},
	title = {GitHub typo corpus: A large-scale multilingual dataset of misspellings and grammatical errors},
	year = {2019}}

@misc{github:jsonrepair,
	author = {Jos de Jong},
	howpublished = {\url{https://jsoneditoronline.org/indepth/parse/fix-json/}},
	note = {Accessed: 2025-09-11},
	title = {JSON Repair: How to fix JSON and validate it with ease},
	year = {2022}}

@article{lee2024bug,
	author = {Lee, Hokyung and Sharma, Sumanyu and Hu, Bing},
	journal = {arXiv preprint arXiv:2406.15325},
	title = {Bug in the code stack: Can llms find bugs in large python code stacks},
	year = {2024}}

@book{fellbaum1998wordnet,
	author = {Fellbaum, Christiane},
	publisher = {MIT press},
	title = {WordNet: An electronic lexical database},
	year = {1998}}

@misc{huggingface:arxiv_abstracts_dataset,
	author = {CShorten},
	howpublished = {\url{https://huggingface.co/datasets/CShorten/Last-Week-on-ML-ArXiv}},
	note = {Accessed: 2025-09-11},
	title = {arXiv Dataset: Hugging Face Dataset Card},
	year = {2022}}

@inproceedings{silveira14gold,
	author = {Natalia Silveira and Timothy Dozat and Marie-Catherine de Marneffe and Samuel Bowman and Miriam Connor and John Bauer and Christopher D. Manning},
	booktitle = {Proceedings of the Ninth International Conference on Language Resources and Evaluation (LREC-2014)},
	title = {A Gold Standard Dependency Corpus for {E}nglish},
	year = {2014}}

@article{mitton1980birkbeck,
	author = {Mitton, Roger and others},
	journal = {Oxford Text Archive Legacy Collection},
	publisher = {University of Oxford},
	title = {Birkbeck spelling error corpus},
	year = {1980}}

@inproceedings{Zhang2015CharacterlevelCN,
	address = {Cambridge, MA, USA},
	author = {Zhang, Xiang and Zhao, Junbo and LeCun, Yann},
	booktitle = {Proceedings of the 29th International Conference on Neural Information Processing Systems - Volume 1},
	location = {Montreal, Canada},
	numpages = {9},
	pages = {649--657},
	publisher = {MIT Press},
	series = {NIPS'15},
	title = {Character-level convolutional networks for text classification},
	year = {2015}}

@article{austin2021program,
	author = {Austin, Jacob and Odena, Augustus and Nye, Maxwell and Bosma, Maarten and Michalewski, Henryk and Dohan, David and Jiang, Ellen and Cai, Carrie and Terry, Michael and Le, Quoc and others},
	journal = {arXiv preprint arXiv:2108.07732},
	title = {Program synthesis with large language models},
	year = {2021}}

@inproceedings{jain2025livecodebench,
	author = {Jain, Naman and Gu, Alex and Li, Wen-Ding and Yan, Fanjia and Zhang, Tianjun and Wang, Sida and Solar-Lezama, Armando and Sen, Koushik and Stoica, Ion},
	booktitle = {International Conference on Learning Representations},
	pages = {58791--58831},
	title = {Livecodebench: Holistic and contamination free evaluation of large language models for code},
	volume = {2025},
	doi = {https://doi.org/10.48550/arXiv.2403.07974},
	year = {2025}}

@article{llama3modelcard,
	author = {AI@Meta},
	title = {Llama 3 Model Card},
	url = {https://github.com/meta-llama/llama3/blob/main/MODEL_CARD.md},
	year = {2024}}

@article{gemma_2024,
	author = {Gemma Team},
	doi = {10.34740/KAGGLE/M/3301},
	publisher = {Kaggle},
	title = {Gemma},
	url = {https://www.kaggle.com/m/3301},
	year = {2024}}

@misc{qwen3technicalreport,
      title={Qwen3 Technical Report}, 
      author={Qwen Team},
      year={2025},
      eprint={2505.09388},
      archivePrefix={arXiv},
      primaryClass={cs.CL},
      url={https://arxiv.org/abs/2505.09388}, 
}

@article{Pedregosa2011tk,
	author = {Pedregosa, F. and Varoquaux, G. and Gramfort, A. and Michel, V. and Thirion, B. and Grisel, O. and Blondel, M. and Prettenhofer, P. and Weiss, R. and Dubourg, V. and Vanderplas, J. and Passos, A. and Cournapeau, D. and Brucher, M. and Perrot, M. and Duchesnay, E.},
	journal = {Journal of Machine Learning Research},
	pages = {2825--2830},
	title = {Scikit-learn: Machine Learning in {P}ython},
	doi = {10.5555/1953048.2078195},
	volume = {12},
	year = {2011}}

@misc{nostalgebraist2020logitlens,
  author       = {nostalgebraist},
  title        = {interpreting {GPT}: the logit lens},
  howpublished = {LessWrong},
  year         = {2020},
  url          = {https://www.lesswrong.com/posts/AcKRB8wDpdaN6v6ru/interpreting-gpt-the-logit-lens}
}

@misc{neuronpedia_jacobian_lens,
  author       = {{Neuronpedia}},
  title        = {{neuronpedia/jacobian-lens}},
  howpublished = {\url{https://huggingface.co/neuronpedia/jacobian-lens}},
}

@misc{openai_gpt4omini_docs,
  author       = {{OpenAI}},
  title        = {{GPT-4o Mini Model}},
  howpublished = {\url{https://developers.openai.com/api/docs/models/gpt-4o-mini}},
}

@misc{anthropic2025haiku45,
  author       = {{Anthropic}},
  title        = {{Introducing Claude Haiku 4.5}},
  year         = {2025},
  month        = oct,
  howpublished = {\url{https://www.anthropic.com/news/claude-haiku-4-5}},
}

@misc{google_gemini_35_flash_lite_docs,
  author       = {{Google}},
  title        = {{Gemini 3.5 Flash Lite Model}},
  howpublished = {\url{https://ai.google.dev/gemini-api/docs/models/gemini-3.5-flash-lite}},
}

\end{document}